\documentclass[article]{article}

\pdfoutput=1
\usepackage{graphicx}
\usepackage{amsmath,amssymb}
\usepackage{bm}
\usepackage{caption}
\usepackage{subcaption}
\usepackage{parskip}
\usepackage{float}
\usepackage{mathrsfs} 
\usepackage{verbatim}
\usepackage[sort&compress,numbers]{natbib}
\usepackage[title]{appendix}

\usepackage{hyperref}

\begin{document}

\title{Influence of non-hydrodynamic forces on the elastic response of an ultra-thin soft coating under fluid-mediated dynamic loading}

\author{Pratyaksh Karan, Jeevanjyoti Chakraborty, Suman Chakraborty \\ Department of Mechanical Engineering, \\ Indian Institute of Technology Kharagpur}
\date{}

\maketitle

\begin{abstract}
The force between two approaching solids in a liquid medium becomes increasingly large with decreasing separation, a phenomenon that prevents contact between the two solids. This growth in force occurs because of the intervening liquid, and, studies of such physical systems constitute the classical discipline of lubrication. Furthermore, when the solid(s) are soft, there are quantitative as well as qualitative alterations in the force interaction due to the solids' deformation. The underlying physics as well as resultant system behaviour are even more complex when forces of non-hydrodynamic origin come into play, two major classes of such forces being the DLVO (Derjaguin-Landau-Verwey-Overbeek) forces and the non-DLVO molecular forces. Studies assessing the coupling of these physical phenomenon are avenues of contemporary research. With this view, we perform an analytical study of fluid-mediated oscillatory motion of a rigid sphere over an ultra-thin soft coating, delineating the distinctive effects of solvation force as well as substrate compliance. Our key finding is the major augmentation in the force and substrate-deformation characteristics of the system due to solvation force when the confinement reduces to a few nanometers. Consideration of solvation force leads to upto four orders of magnitude and upto three orders of magnitude increment in force and substrate-deformation respectively. While higher softness leads to higher deformation (as expected), its effect on force and substrate-deformation characteristics exhibits a tendency towards amelioration of the increment due to solvation force.
\end{abstract}

\section{Introduction} \label{sec:intro}
Understanding the interaction between a fluid and its confining solid boundaries is immensely important in a wide variety of applications ranging from modelling of human physiology to industrial and scientific tribological applications \cite{Williams2005}. The motion of an object close to a wall with intervening fluid is one of the cornerstones of lubrication studies, where an established finding is the growth in lift generated on the object upon approach, enabling the design of robust and effective bearing setups \cite{Cooley1969,Chan1985,Hamrock2004}. Furthermore, there is rich literature on `soft lubrication' too, where either the approaching object or the wall or both are deformable \cite{Higginson1962,Dowson1995,Yin2005,Mahadevan2005,Chu2006,Chu2006a,Shen2007,Shinkarenko2009,Shinkarenko2009a,Chakraborty2010,Balmforth2010,Chakraborty2011,Snoeijer2013,Scaraggi2014,Dong2015,Stupkiewicz2016,Pandey2016,Karan2018,Zhang2019,Zhao2019}, which finds applications in topics ranging from biotransport modelling \cite{Weekley2006,Jones2008,Butler2008,Davies2018} to  tool design and analysis\cite{Larsson1997,Dowson1999,Li2010}. Such soft-lubrication setups are wide-spread in both natural and man-made world, examples being scanning probe microscope (SPM) and surface force apparatus (SFA) setups \cite{Restagno2002,Jones2005,Butt2005,Leroy2011,Leroy2012,Villey2013,Carpentier2015,Wang2015,Wang2017,Wang2017a} and motion of biological entities like red blood corpuscles (RBCs) in fluidic environments \cite{Sukumaran2001,Beaucourt2004,Trouilloud2008}.\\
With classical purely-hydrodynamic studies in hard- and soft-lubrication \cite{Brenner1961,Stewartson1967,Goldman1967,Goldman1967a,Hinch1986} serving as a solid foundation for further explorations, a number of later studies considered the additional effects of electrokinetics and van der Waals forces  \cite{Davis1986,Bike1990,VandeVen1993,VandeVen1993a,Prieve1995a,Prieve1995,Wu1996,Warszynski1998,Tabatabaei2006,Tabatabaei2006a,Chakraborty2010,Urzay2010}. An insightful finding of these works is that the presence of an electrical double layer (EDL) leads to increase in the lift on the object. On the other hand, effects of van der Waals force are not as straightforward and exhibit dependence on the interplay of substrate softness and Hamaker's constant. Overall, these studies have expanded the scope of soft-lubrication through the incorporation of forces of non-hydrodynamic origin. More specifically, the force interactions studied in these works include EDL disjoining and van der Waals forces, which together are termed as DLVO (Derjaguin-Landau-Verwey-Overbeek) forces, and fall under the purview of continuum description of the intervening fluid in regards to their theoretical modelling. \\
Consideration of DLVO forces assists in satisfactory modelling of `object-near-a-wall' setups with object-wall separation in the range of 10-100 nm. However, for similar setups with smaller object-wall separations, considering only the hydrodynamic and DLVO force interactions becomes inadequate in explaining observed phenomena \cite{Pashley1984,Israelachvili1987,Israelachvili1988}. For such systems, non-continuum forces of molecular interactions between various materials present, i.e., non-DLVO molecular forces, become increasingly important \cite{Israelachvili1987,Henderson2000}. Factoring in these forces apart from hydrodynamic and DLVO forces is crucial for understanding soft-lubrication at nanometric separations given the fact that at such separations, these forces heavily dominate the force interactions between the object and the wall. While the experimental and statistical mechanics studies of non-DLVO forces span a few decaded \cite{Henderson1976,Snook1978,Snook1980,Pashley1981,Pashley1981a,Pashley1982,Israelachvili1982,Pashley1984,Israelachvili1988,Gao1997,Henderson2000,Qin2003,Zhang2005,Gao2007,Israelachvili2011,Yang2011}, there is an evident drift of the academic community to their inclusion in detailed theoretical soft-lubrication \cite{Jang1995,Matsuoka1996,Matsuoka1997,AlSamieh2001,Zhang2019a}. \\
Hence, with the intent of contributing to the growing research literature on incorporating these force interactions for nano-scale tribological studies, an analytical treatment of oscillatory motion of a rigid sphere over a soft ultra-thin substrate coated on a rigid platform with intervening fluid is presented in this article. The emphasis is on studying the role of solvation force, i.e. the force between surfaces due to confinement-induced structuring of intervening fluid \cite{Wasan2001,Israelachvili2011}, which has been incorporated as a closed-form expression. For the setups studied, hydrodynamic pressure is found to be much smaller than DLVO and solvation forces, and thus, the latter dominate the system behaviour throughout. Furthermore, EDL disjoining pressure dominates for relatively larger separations while van der Waals and solvation forces dominate for smaller separations, with solvation force emerging as the dominant force at extremely small separations, rendering hydrodynamic and DLVO forces negligible. The consideration of solvation force along with hydrodynamic and DLVO forces accounts for amplification of upto four orders of magnitude in force between the surfaces, and upto three orders of magnitude in substrate deformation. This study essentially assimilates the non-continuum solvation force into the continuum-description based mathematical framework for a specific soft-lubrication setup, viz., rigid sphere oscillating above soft ultra-thin coating. Therefore, it serves as an extended pseudo-continuum model capable of accounting for the effects of high-proximity-induced non-hydrodynamic continuum and molecular force interactions in determining the behaviour of similar physical systems. Our model contributes to a growing family of such pseudo-continuum models that constitute a currently-maturing framework capable of complementing the computationally expensive molecular dynamic simulations for related problems, where an in-depth analysis of underlying molecular physics is not required.\\
\begin{figure}[h!]
\centering
\includegraphics[height=8cm]{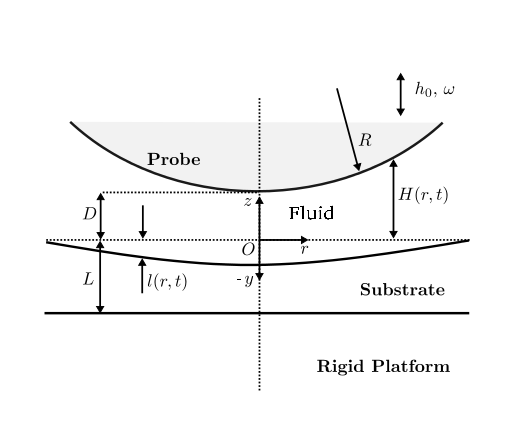}
\caption{Schematic of an oscillating rigid sphere near a ultra-thin soft coating}
\label{fig:schematic}
\end{figure}
\section{Problem Description}\label{sec:prob}
Sinusoidal oscillations of a rigid sphere close to a soft thin substrate layer coated on a rigid platform with intervening fluid along an axis perpendicular to the undeformed fluid-substrate interface is considered (as presented in figure \ref{fig:schematic}). The major system characteristics of concern are the total force between the surfaces and the fluid-substrate interface deflection (henceforth referred to as `force' and `deflection' respectively).
\subsection{Model Setup}\label{subsec:model}
The substrate material is modelled as a linear-elastic and isotropic solid, with a flat horizontal profile for the undeformed fluid-substrate interface. The intervening fluid is a dilute aqueous monovalent binary electrolytic solution. The characteristic undeformed gap height ($D$) as well as the undeformed substrate thickness ($L$) are small compared to the sphere radius throughout the cases studied. The co-ordinate system used and the pertinent length scales of the setup are presented in figure \ref{fig:schematic}, with $\displaystyle r\text{-}z$ notation used for all variables pertaining to fluid flow and $\displaystyle r\text{-}y$ notation for all variables pertaining to substrate deformation ($y$ is positive in the upward direction). \\
The substrate deformation components are denoted by $\displaystyle u$, with the appropriate subscripts representing the direction of the component. Total pressure is denoted by $\displaystyle p$, which is the sum of hydrodynamic and non-hydrodynamic pressure components. Hydrodynamic (hd) pressure is denoted by $\displaystyle p^{\star}$, while non-hydrodynamic pressure components are denoted by $\displaystyle \Pi$ with the subscripts of $\displaystyle \text{EDL}$, $\displaystyle \text{vdW}$ and $\displaystyle \text{sol}$ for EDL disjoining pressure, van der Waals pressure and solvation pressure respectively. The oscillation amplitude and frequency of the sphere are $\displaystyle h_{0}$ and $\displaystyle \omega$ respectively, and $\displaystyle D$ is the mean separation of the sphere from the origin. The substrate is stuck to a rigid platform at the bottom. The deflection $\displaystyle l$ is positive when opposite to the direction of $\displaystyle y$-axis, hence the gap height is $\displaystyle H+l$. In the lubrication region, the expression for $\displaystyle H$ is, \\
\begin{equation}
H = D + \frac{r^2}{2R} + h_{0} \cos(\omega t).
\label{eq:sphere}
\end{equation}
The deflection profile $\displaystyle l$ is part of the solution (given by $\displaystyle l(r,t) = -u_{y}(r,y,t)$ at $\displaystyle y~=~0$).\\
The characteristic length, velocity, pressure, time and deformation scales are presented on the right side of table \ref{tab:nondim}, where the length, velocity and pressure scales are in tandem with classical hydrodynamic studies of lubrication setups. The time scale is taken as inverse of the sphere's oscillation frequency. The characteristic scale of substrate deformation ($\displaystyle \lambda$) is obtained from the substrate-fluid traction balance condition and is specified in subsection \ref{subsec:molsemi}. The pertinent length scale ratios for the mathematical formulation are presented in the left side of table \ref{tab:nondim}. A crucial facet of the setup being studied is that the oscillation amplitude of the sphere is so close to the characteristic undeformed gap height ($D$) that the length-scale along $z$-axis varies substantially throughout the oscillation and is therefore dependent on time. As a result, the $z$-axis length scale has to be considered separately for each time instant and hence, all the associated non-dimensional parameters and governing equation co-efficients become time-dependent as well. Although this approach deviates from scaling conventions, it is anticipated that such time-dependent scaling would not yield incorrect results so long as the simplifications to the pertinent governing equations and boundary conditions remain consistent for all time-instants in the duration of the study. Hence, the characteristic length-scale for $z$-axis, $d(t)$, is,
\begin{equation}\label{dt}
d(t) = D + h_{0}\cos(\omega t) = D(1+\alpha\cos(\omega t)).
\end{equation}
\subsection{Mathematical Formulation}\label{subsec:gdes}
The setup studied is mathematically represented by the continuity and momentum conservation equations for the fluid flow and mechanical equilibrium equation for the sustrate deformation, along with the no-slip and no-penetration boundary conditions at the fluid-substrate and the fluid-sphere interfaces closing the fluid flow problem and no-deformation at substrate-platform interface and traction-balance condition at fluid-substrate interface closing the substrate deformation problem. The equations and boundary conditions are non-dimensionalized with the characteristic scales as presented in table \ref{tab:nondim}, and expressions presented henceforth are non-dimensionalized. The system is evidently axisymmetric and non-rotating. The complete set of non-dimensionalized equations and boundary conditions are presented in Appendix \ref{sec:ap_gdes}.\\
Following the traditional simplifications and subsequent analysis of soft lubrication studies \cite{Mahadevan2004,Glover2007,Chakraborty2010,Urzay2010,Chakraborty2011}, we obtain the Reynolds equation,
\begin{equation}
\frac{\epsilon}{\epsilon_{0}}\frac{\eta}{\alpha}\frac{\partial l}{\partial t}-\sin(t) = \frac{1}{12r} \frac{\partial}{\partial r} \left[r  (H+\eta l)^3 \frac{\partial p^{\star}}{\partial r}\right], \label{eq:Reeq}
\end{equation}
which is subject to the boundary conditions,
\begin{equation}
\label{eq:pbc}
p^{\star} = 0~~~~~~\text{as}~~~r \rightarrow \infty,
\end{equation}
\begin{equation}
\label{eq:pbc1}
\frac{\partial p^{\star}}{\partial r} = 0~~~\text{at}~~~r = 0,
\end{equation}
and expression for substrate deformation,
\begin{equation}
\label{eq:uy}
u_{y} = -\frac{\mu\omega\alpha\epsilon_{0}}{\epsilon^2\kappa E_{y}}\frac{(1+\nu)(1-2\nu)}{(1-\nu)}(1+y)p,
\end{equation}
leading to the expression for $l$ (which simply equals $- u_{y}$ at $y=0$) as,
\begin{equation}
\label{eq:l}
l = \frac{\mu\omega\alpha\epsilon_{0}}{\epsilon^2\kappa E_{y}}\frac{(1+\nu)(1-2\nu)}{(1-\nu)}p,
\end{equation}
where $E_{y}$ is the substrate's Young's modulus and $\nu$ is its Poisson's ratio. It should be noted that as the substrate material behaviour approaches incompressibility, the formulation and scales vary significantly from the one that has been employed to derive equation \eqref{eq:l}. Therefore, this expression is valid only for compressible substrates and doesn't stand applicable for incompressible substrates. A brief discussion and scaling analysis for incompressible substrates is presented in Appendix \ref{sec:ap_incompr}.\\
The pressure in expression \eqref{eq:l} comprises not only the hydrodynamic pressure but also the EDL disjoining, van der Waals and solvation pressure components. Thus, $p$ is the total pressure,
\begin{equation}
\label{eq:ptot}
p = p^{\star}+\Pi_{\text{EDL}}+\Pi_{\text{vdW}}+\Pi_{\text{sol}}.
\end{equation}
The comparison between sphere-origin separation and sphere-substrate interface deflection depends on both the elastic properties of the substrate material as well as non-hydrodynamic pressure characteristics, and could lead to `one-sided' dependence of substrate deformation on the flow dynamics (when deflection is negligible compared to sphere-origin separation) or `two-sided' coupling (when deflection is comparable to sphere-origin separation).
\begin{table}[h!]
\small
\caption{Assigned Notations of Length Scale Ratios and Characteristic Values of System Variables (time dependent ratios and characteristic values demarcated by attaching `(t)'); $\mu$ is the fluid viscosity}
\label{tab:nondim}
\begin{tabular*}{1.0\textwidth}{@{\extracolsep{\fill}}ccccccc}
\hline
\textbf{Length Scale}  									& 	\textbf{Assigned}			&~&~&~&	\textbf{System} 		& 	\textbf{Characteristic} 									\\
\textbf{Ratio}											&	\textbf{Notation}			&~&~&~&	\textbf{Variable}		&	\textbf{Value}												\\
\hline \\[3pt]
$\displaystyle \frac{D}{R}$   							& 	$\displaystyle \epsilon_{0}$&~&~&~&	$\displaystyle r$ 		& 	$\displaystyle \epsilon(t)^\frac{1}{2} R$															\\[10pt]
$\displaystyle \frac{d(t)}{R}$   						& 	$\displaystyle \epsilon(t)$	&~&~&~&	$\displaystyle z, H$	& 	$\displaystyle d(t)$														\\[10pt]
$\displaystyle \frac{h_{0}}{D}$   						& 	$\displaystyle \alpha$ 		&~&~&~&	$\displaystyle y$ 		& 	$\displaystyle L$															\\[10pt]
$\displaystyle \frac{L}{R}$  							& 	$\displaystyle \delta$ 		&~&~&~&	$\displaystyle v_{r}$ 	& 	$\displaystyle \frac{\omega h_{0}}{\epsilon(t)^\frac{1}{2}}$ 							\\[10pt]
$\displaystyle \frac{\lambda(t)}{L}$ 					& 	$\displaystyle \kappa(t)$ 	&~&~&~&	$\displaystyle v_{z}$ 	& 	$\displaystyle \omega h_{0}$														\\[10pt]
$\displaystyle \frac{\kappa(t)\delta}{\epsilon(t)}$		& 	$\displaystyle \eta(t)$		&~&~&~&	$\displaystyle t$		& 	$\displaystyle \frac{1}{\omega}$											\\[10pt]
$\displaystyle \frac{\delta}{\epsilon(t)^\frac{1}{2}}$	& 	$\displaystyle \gamma(t)$	&~&~&~&	$\displaystyle p$		&	$\displaystyle \frac{\mu\omega\alpha\epsilon_{0}}{\epsilon(t)^{2}}$		\\[10pt]
~														&	~ 							&~&~&~&	$\displaystyle u, l$	&	$\displaystyle \lambda(t)$ 												\\
\hline
\end{tabular*}
\end{table}
\subsection{Non-Hydrodynamic Pressure Components}
While hydrodynamic pressure directly affects the flow dynamics, the non-hydrodynamic pressure components indirectly affect the flow dynamics by altering deflection, which appears in the Reynolds equation. Furthermore, the implications of non-hydrodynamic pressure components on deformation and force characteristics of the system are crucial and exclusive. \\
Of the pressure components in expression \eqref{eq:ptot}, EDL disjoining and van der Waals pressure components constitute the DLVO forces. The EDL disjoining pressure is the osmotic pressure generated by the non-uniform distribution of ionic species (occurring from the interplay between entropic and electrostatic effects) when the intervening fluid is an electrolytic solution. This osmotic pressure becomes significant as the EDL overlap increases and it is approximately represented by the exponentally decaying expression \cite{Russel1991},
\begin{equation}
\Pi_{\text{EDL}} = \frac{64n_{0}k_{\text{B}}T\tilde{\zeta}^2\epsilon^2}{\mu\omega\alpha\epsilon_{0}} \ \exp \left(-\epsilon K R (H+\eta l) \right), \label{eq:piDL}
\end{equation}
where $\tilde{\zeta}$ is $\displaystyle \tanh\left(\frac{q\zeta}{4k_{\text{B}}T}\right)$, $\displaystyle K = \sqrt{\frac{2q^2n_{0}}{\bar{\epsilon}\bar{\epsilon}_{0}k_{\text{B}}T}}$ is the inverse Debye length, $\displaystyle \zeta$ is the zeta-potential, $\bar{\epsilon}$ is the relative permittivity of the fluid, $\bar{\epsilon}_{0}$ is the permittivity of vacuum, $k_{\text{B}}$ is Boltzmann constant, $q$ is elementary charge, and $n_{0}$ is the salt's electroneutral number density. On the other hand, van der Waals force is a force between surfaces arising out of the aggregate electrostatic interactions between the induced dipoles on the surfaces, their respective bulks and the intervening fluid. The van der Waals pressure between two large surfaces (in comparison to the separation between them), subject to certain conditions regarding the distance between the surfaces \cite{Tabor1969,Israelachvili1978,Israelachvili2011}, varies inversely with the cube of the separation between the surfaces and grows to significant magnitude at small separations. Its expression is,
\begin{equation}
\Pi_{\text{vdW}} = -\frac{A_{\text{sfw}}}{6\pi\epsilon\epsilon_{0}\alpha\mu\omega R^3} \frac{1}{(H+\eta l)^3}, \label{eq:pivdW}
\end{equation}
where, $\displaystyle A_{\text{sfw}}$ is the Hamaker's constant which is typically of the order of $10^{-20}$ Joules \cite{Israelachvili2011}. It should be noted that the DLVO forces are often expressed in the form of interaction energy between two bodies, with the force between bodies being the derivative of the interaction energy with separation. Furthermore, both the DLVO force expressions for a sphere-surface pair transform into the DLVO force (per unit area) expressions for surface-surface pair for separations much smaller than the sphere's radius, yielding the pressure expressions \eqref{eq:piDL} and \eqref{eq:pivdW} \cite{Israelachvili2011}.\\
The force between surfaces at separations smaller than $\sim$ 5 nm include non-trivial components arising out of the molecular nature of the intervening liquid. These forces become significant and much larger in comparison to DLVO forces at length scales where DLVO theory becomes inapplicable \cite{Mitchell1978,Israelachvili1982b,Boinovich2007}. For any system consisting of a fluid-solid interface, fluid molecules close to the interface structure themselves in ordered layers with the ordering dampening out into the fluid bulk in about five to ten times the fluid particle size \cite{Tarazona1985,Mezic1998}. Thus, confinement of the fluid molecules by surfaces separated by lower than five to ten times the fluid particle size causes the structuring of the fluid molecules to oscillate between optimal and pessimal as the separation between the surfaces is varied. The energy variation associated with this variation of packing with separation manifests itself as a force between the two surfaces called the solvation force \cite{Christenson1983,Wasan2001,Israelachvili2011}. This force has a damped oscillatory variation with the surface separation for perfectly smooth surfaces that do not interact with the fluid molecules. It should be noted that force between surfaces in real-world scenarios comprises other effects as well, like fluid-surface energy-interactions (solvophobicity), hydration effects, surface roughness, polymeric steric effects, etc. \cite{Israelachvili1982,Pashley1982,Frink1998,Qin2003,Valle2005,Yang2011,Israelachvili2011} and the applicable short-range force law for any real-world system will involve an assimilation of all force interactions specific to that system. However, in the current study, we focus on assessing the effects of the resultant solvation force due to short-range fluid structuring for the system being studied and assimilation of other short-range energy interactions will be done in future studies. Focusing on the mathematical expression for solvation pressure, while a theoretically rigorous and reliable closed-form expression for simple water-like fluids is a topic of active research, studies of solvation pressure in available literature indicate that it has a damped-oscillatory profile with exponential damping and sinusoidal oscillations and the decay as well as oscillation length for simple spherical fluids is approximately the fluid particle size. Therefore, we consider the expression for solvation pressure as,
\begin{equation}\label{eq:piS}
\Pi_{\text{sol}} = \frac{\Lambda\epsilon^2}{\mu\omega\alpha\epsilon_{0}}\exp\left(-\frac{\epsilon R (H+\eta l)}{s}\right) \cos\left(\frac{2\pi\epsilon R (H+\eta l)}{s}+\phi\right)
\end{equation}
where, $\Lambda$ is the solvation pressure amplitude, $\displaystyle s$ is the fluid particle size, and $\phi$ is the solvation pressure phase. A brief summarization of our literature survey on solvation forces is presented in Appendix \ref{sec:ap_piS}.
\section{Solution}\label{sec:soln}
Since the system behaviour is quasi-steady for flow dynamics and quasi-static for substrate deformation, the solution for one complete oscillation is representative of the complete solution for any particular set of parameters. Hence, solution methodology for one complete oscillation has been formulated.\\
From the mathematical modelling and associates simplification of the governing equations and boundary conditions, equations \eqref{eq:Reeq} and \eqref{eq:l} emerge as the coupled governing equations, subjected to boundary conditions \eqref{eq:pbc} and \eqref{eq:pbc1} and conjugated by expressions \eqref{eq:piDL}, \eqref{eq:pivdW} and \eqref{eq:piS}. A straightforward analytical solution of the problem is evidently not plausible, and therefore asymptotic and semi-analytical methodologies are posited, as presented ahead.
\subsection{Classic Asymptotic Approach}\label{subsec:classpert}
Proceeding with the typical classical asymptotic solution approach to such soft lubrication problems that is being studied \cite{Glover2007,Urzay2010}, $\eta$ is taken as the small parameter for the perturbation expansion. Although the value of $\eta$ is unknown, the asymptotic solution methodology is presented here and discussion on the expression for $\eta$ is presented in subsection \ref{subsec:molsemi}, along with discussion on conditions where classical approach doesn't hold good. Hence, continuing with $\eta$ as the small parameter, the perturbation expansions of relavent variables are,
\begin{equation}
\label{eq:pertAp}
p_{i} = p_{i(0)} + \eta p_{i(1)} + O(\eta^2),
\end{equation}
\begin{equation}
\label{eq:pertAl}
l = l_{(0)} + \eta l_{(1)} + O(\eta^2).
\end{equation}
The leading order solution for hydrodynamic pressure is obtained as,
\begin{equation}
\label{eq:p0st}
p^{\star}_{(0)} = \frac{3\sin(t)}{H^2},
\end{equation}
and the leading order solutions of the non-hydrodynamic pressure components are simply their expressions (i.e. expressions \eqref{eq:piDL}, \eqref{eq:pivdW}, \eqref{eq:piS}) without the deflection term. The first order solution of hydrodynamic pressure is obtained by numerically solving the first order split of the Reynolds equation. The leading and first order solutions of deflection are simply the split of expression \eqref{eq:l} for each order.
\subsection{Semi-analytical Approach}\label{subsec:molsemi}
We first obtain an expression for $\eta$, so as to facilitate the utilization of the method in subsection \ref{subsec:classpert} as well as ascertain its limitations. Focussing on the expression for deflection \eqref{eq:l}, it can be interpreted that deflection is the effect and the total pressure is the cause. Therefore, in keeping with the conventions of scaling analysis, the non-dimensionalized deflection and total pressure terms should scale equal, and hence, the factor multiplied to total pressure should scale as 1, which gives the expression for $\kappa$. With $\kappa$ being the only unknown term in the expression for $\eta$ (which is $\displaystyle \frac{\kappa\delta}{\epsilon}$), the expression for $\eta$ is obtained as, 
\begin{equation}
\eta = \frac{\mu\omega\alpha\epsilon_{0}\delta}{\epsilon^3 E_{y}}\frac{(1+\nu)(1-2\nu)}{(1-\nu)}. \label{eq:kappa}
\end{equation}
The expression for $\eta$ indicates that it depends on the fluid and substrate material properties ($\mu$, $E_{y}$ and $\nu$) as well as the system dimensions and imposed dynamics ($\omega$, $\alpha$, $\epsilon_{0}$, $\delta$ and $\epsilon$). Furthermore, the magnitude of $\eta$ varies with time due to its dependence of $\epsilon$. Therefore, $\eta$ is an imposed time-dependent parameter of the system. Furthermore, getting the expression for $\eta$ is tantamount to getting the expression for $\lambda$ since $\displaystyle \lambda = \kappa L = \frac{\epsilon\eta}{\delta}L$. \\ 
Two conditions needs to be considered when the asymptotic method presented in subsection \ref{subsec:classpert} would be rendered inapplicable. The first condition emerges from the fact that hydrodynamic pressure and all non-hydrodynamic pressure components are non-dimensionalized with the classical lubrication hydrodynamic pressure scale. However, at very small sphere-substrate separations, the non-hydrodynamic pressure components exceed this scale. Therefore, the non-dimensionalized total pressure and resultantly the deflection will exceed unity by approximately a magnification factor $M$,
\begin{multline}
M = 1+\frac{\epsilon^{2}}{\mu\omega\alpha\epsilon_{0}}\left[64n_{0}k_{\text{B}}T\tilde{\zeta}^2\exp\left(-\epsilon KR\right)+  \right.\\ \left.
\frac{A_{\text{sfw}}}{6\pi\epsilon^3 R^{3}}+\Lambda\exp\left(-\frac{\epsilon R}{s}\right)\right].
\end{multline}
For the asymptotic method presented in subsection \ref{subsec:classpert} to hold good, it is required that the leading order deflection term (i.e. $\eta l_{(0)}$) scale at least an order smaller than $H$, which scales as 1. Therefore, if $\eta M > 0.1$, the asymptotic method ceases to be applicable. The second condition comes from the restriction that the first order expressions for non-hydrodynamic pressure components as per the perturbation split of subsection \ref{subsec:classpert} do not exceed the leading order solution, the violation of which would be tantamount to divergence of the perturbation scheme. From the first order expressions of all three non-hydrodynamic pressure components, a divergence factor $N$ is obtained as,
\begin{multline}
N = \max\left[\eta\epsilon KRM, 3\eta M, \right. \\ \left. \frac{\eta\epsilon RM}{s}\left(1+2\pi\tan\left(\frac{2\pi\epsilon R}{s}\left(1+\eta M\right)+\phi_{1}\right)\right) \right].
\end{multline}
If $N > 1$, the asymptotic method ceases to be applicable. However, this check needs to be applied only when the non-hydrodynamic pressure components come into significance in comparison to hydrodynamic pressure, a reasonable condition for which can be taken as $M>0.01$. Therefore, when $\eta M$ exceeds 0.1 or $N$ exceeds 0.1 given $\eta M$ is higher than 0.01 or both, then the asymptotic method has to be replaced by an iterative numerical approach which is discussed ahead. \\
The deflection is given by expression \eqref{eq:l}, with the non-hydrodynamic pressure component expressions being \eqref{eq:piDL}, \eqref{eq:pivdW} and \eqref{eq:piS} and hydrodynamic pressure being the solution of equation \eqref{eq:Reeq}, subject to boundary conditions \eqref{eq:pbc} and \eqref{eq:pbc1}. This leads to expression \eqref{eq:l} becoming an implicit equation in $l$,
\begin{equation}
\label{eq:limp}
l = \frac{\mu\omega\alpha\epsilon_{0}}{\epsilon^2\kappa E_{y}}\frac{(1+\nu)(1-2\nu)}{(1-\nu)}p(l) = p(l) = \mathscr{F}\left(l\right),
\end{equation}
which is solved numerically at any time-instant and any point on the $r$-axis using the iterative bisection root-finding method where the solution of the last time step is used as the guess. While the expressions for non-hydrodynamic pressure components are available as closed-form functions of $l$, same is not true for hydrodynamic pressure. However, since hydrodynamic pressure becomes small in comparison to non-hydrodynamic pressure components when the semi-analytical method is employed, the hydrodynamic pressure is numerically solved for with iterations of the finite-difference solution of Reynolds equation enveloped over all the node-wise solutions of $l$. \\
From the solution methodology, $\eta$, $M$ and $N$ emerge crucial in determining the solution method as well as in providing insights on the system behaviour. The first parameter, $\eta$, that compares deflection to sphere-origin separation $d(t)$, represents the effect of deflection on the flow dynamics (by altering the flow bounds) and is dependent on fluid and solid properties and the imposed system geometry and dynamics. The second parameter, $M$, highlights the growth of non-hydrodynamic pressure components to dominate over hydrodynamic pressure, and is dependent on all the system properties on which $\eta$ is dependent as well as parameters corresponding to non-hydrodynamic pressure components, i.e. fluid particle size, solvation pressure amplitude and phase, Hamaker's constant, surface zeta-potential, fluid dielectric constant and electrolyte concentration. The third parameter, $N$ keeps a check on the divergence of the non-hydrodynamic pressure components pertaining to the asymptotic method, depends on the same system properties that $M$ is dependent upon, and is primarily a mathematical parameter and doesn't provide any major insights about the system behavior that can't already be drawn from $\eta$ and $M$.
\begin{table}[h!]
\small
\caption{Property/Parameter values used for obtaining representative solution}
\label{tab:paramtaken}
\begin{tabular*}{1.0\textwidth}{@{\extracolsep{\fill}}llll}
\hline
\textbf{Parameter}														&	\textbf{Value}								&
\textbf{Parameter}														&	\textbf{Value}								\\[3pt]
\hline
\textbf{Geometry }														& 	~											&
\textbf{Substrate}						 								&	~											\\[3pt]
$R$ (sphere radius)														&	1 mm				 						&
\textbf{`Hard'}															&	~											\\[3pt]
$D$ (reference gap)														&	50 nm				 						&
$E_{y}$																	&	$90$ GPa									\\[3pt]
$L$ (substr. thick.)													&	50 nm				 						&
$\nu$																	&	$0.20$	 									\\[3pt]
$h_{0}$ (osc. ampl.)													&	10 - 49.5 nm								&
$k_{\text{flex}}$														&	$5$x$10^{-4}$ nm/MPa						\\[3pt]
$\omega$ (osc. freq.)													&	1 Hz										&
$\kappa_{\text{max}}$													&	10$^{-6}$									\\[3pt]
$\alpha$ (osc. amp./ref. gap)											&	$\leq 0.99$									&
$\eta_{\text{max}}$														&	10$^{-4}$									\\[3pt]
$\epsilon_{0}$, $\delta$ (aspect ratios)								&	$5 \text{x} 10^{\text{-}5}$					&
\textbf{`Stiff'}														&	~											\\[3pt]
$\gamma_{\text{max}}$ (in an osc.)										&	0.08										&
$E_{y}$																	&	$700$ MPa									\\[3pt]
$T$ (amb. temp.)														&	298.15 K									&
$\nu$ 																	&	$0.43$	 									\\[3pt]
\textbf{Fluid}															&	~											&
$k_{\text{flex}}$														&	$0.025$ nm/MPa								\\[3pt]
$\rho$ (density)														&	$1000$ kg m$^{-3}$							&
$\kappa_{\text{max}}$ 													&	10$^{-4}$									\\[3pt]
$\mu$ (dyn. visc.)														& 	$0.001$ Pa$\cdot$s							&
$\eta_{\text{max}}$														&	10$^{-2}$									\\[3pt]
\textbf{$\Pi_{\text{EDL}}$ parameters}	 								&	~											&
\textbf{`Compliant'}													&	~											\\[3pt]
$n_{0}$ (electro-neutral conc.)											&	1 mM										&
$E_{y}$																	&	$4.25$ MPa									\\[3pt]
$\zeta$ (zeta-potential)												&	100 mV										&
$\nu$																	&	$0.46$	 									\\[3pt]
$\bar{\epsilon}$ (relative permittivity)								&	80											&
$k_{\text{flex}}$														&	$2.545$ nm/MPa								\\[3pt]
\textbf{$\Pi_{\text{vdW}}$ parameter}									&	~											&
$\kappa_{\text{max}}$													&	10$^{-6}$									\\[3pt]
$A_{\text{sfw}}$ (Hamaker's constant)\footnotemark						&	5 x 10$^{-21}$ J							&
$\eta_{\text{max}}$														&	10$^{-4}$									\\[3pt]
\textbf{$\Pi_{\text{sol}}$ parameters}	 								&	~											&
\textbf{`Soft'}															&	~											\\[3pt]
$\Lambda$ (sol. press. amp.)											&	$1.25$ GPa									&
$E_{y}$																	&	$9.5$ kPa									\\[3pt]
$\sigma$ (water particle dia.)											&	$270$ pm									&
$\nu$																	&	$0.492$	 									\\[3pt]
$\phi$ (sol. press. phase)												&	$0$											&
$k_{\text{flex}}$														&	$247.33$ nm/MPa								\\[3pt]
~																		&												&	
$\kappa_{\text{max}}$													&	10$^{-6}$									\\[3pt]
~																		&												&
$\eta_{\text{max}}$														&	10$^{-4}$									\\[3pt]
\hline	
\end{tabular*}
\end{table}
\section{Results}\label{sec:results}
\begin{figure*}[h!]
\centering
\begin{subfigure}[b]{0.5\textwidth}
\centering
\includegraphics[height=7.00cm]{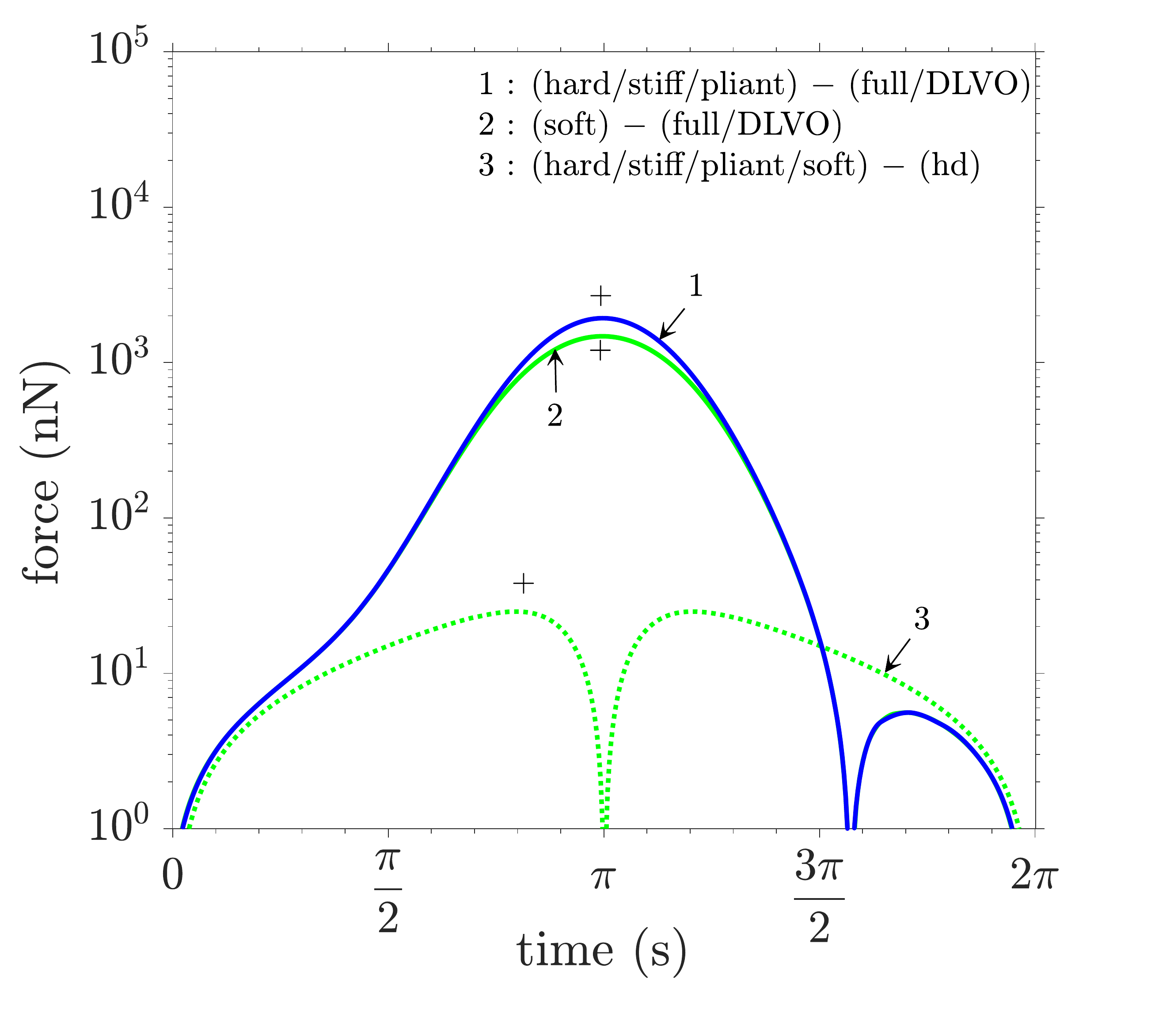}
\caption{\centering}
\label{subfig:F_40d00_pliant_soft}
\end{subfigure}
\begin{subfigure}[b]{0.5\textwidth}
\centering
\includegraphics[height=7.00cm]{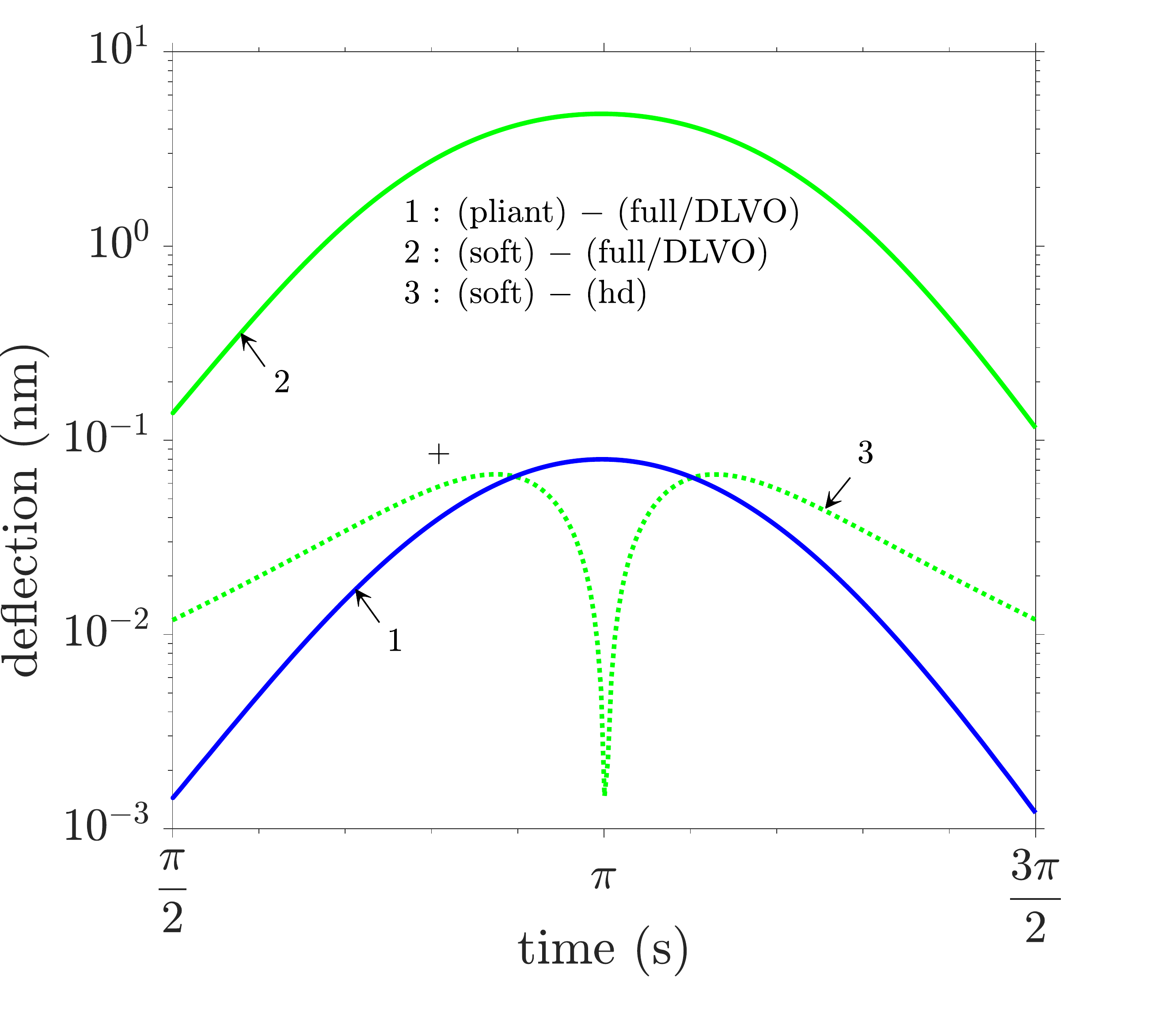}
\caption{\centering}
\label{subfig:lcl_40d00_pliant_soft}
\end{subfigure}
\caption{Evolution of (a) force between sphere and substrate, and, (b) fluid-substrate inteface deflection at the origin, for the 40 nm amplitude case}
\label{fig:40d00}
\end{figure*}

The parameters corresponding to system geometry, imposed dynamics, fluid and solid properties, and DLVO and solvation pressure components are presented in table \ref{tab:paramtaken}. The solvation pressure parameters, i.e., amplitude ($\Lambda$) and phase ($\phi$), are taken from the data for the theoretically-obtained solvation pressure profile between smooth surfaces in a solvent of volume fraction 0.3665 \cite{Wasan2001}. \\
Results corresponding to four \textit{substrates} are obtained. Ranging from hardest to softest, these substrates are labelled `hard', `stiff', `pliant', and `soft'. For each substrate, three \textit{solutions} are obtained - first considering all pressure components and labelled `full', second considering only hydrodynamic and DLVO (i.e. van der Waals and EDL disjoining) pressure components and labelled `DLVO', and third considering only hydrodynamic pressure and labelled `hd'. \\
For each substrate$-$solution combination, \textit{cases} corresponding to amplitudes ranging for 10 nm up to 49.5 nm are studied. The nomenclature of `substrate' to represent softness, `solution' to represent nature of force interactions and `case' to represent amplitude, with the respective definitions and labels, is used consistently throughout the discussion in subsections \ref{subsec:amplitude} and \ref{subsec:characteristics} and appendix \ref{sec:ap_press}. Furthermore, the term `deflection' used in these subsections and appendices refers to the fluid-substrate interface deflection at the origin and the term `force' refers to the total force between the sphere and substrate. Lastly, the plot-line colours grey, magenta, blue and green are used to depict hard, stiff, pliant and soft substrate respectively, and the linestyles unbroken, dashed-dot, and dotted are used to depict full, DLVO and hd solution respectively, in Figures \ref{fig:40d00} to \ref{fig:aggr_rigid_stiff}.\\
The maximum amplitude case studied for the hard and stiff substrates is 49.5 nm. The maximum amplitude case studied for the pliant substrate is 48 nm, because for higher amplitude cases, the system response exhibits adhesive characteristics. The behaviour of non-DLVO forces close to adhesion being complex, we have refrained from going higher than 48 nm. The maximum amplitude case studied for the soft substrate is 40 nm, because for higher amplitude cases, the fluid-substrate interface deflection goes higher than 5 nm, which becomes comparable to the substrate thickness (50 nm) and thus requires treatment with a finite-strain constitutive formulation. \\
Considering the expression for deflection, equation \eqref{eq:l}, the expression relating dimensional deflection and total pressure is,
\begin{equation}\label{eq:deflexp}
l' = \frac{L}{E_{y}}\frac{(1+\nu)(1-2\nu)}{(1-\nu)}p' = k_{\text{flex}}~~p'
\end{equation}
where, $k_{\text{flex}}$ represents the substrate flexibility and has the dimension of length divided by pressure, and the superscript $'$ signifies that the terms are dimensional. Value of $k_{\text{flex}}$ for the four substrates considered is presented in table \ref{tab:paramtaken}.
\subsection{Influence of Oscillation Amplitude}\label{subsec:amplitude}
\begin{figure*}[h!]
\centering
\begin{subfigure}[b]{0.5\textwidth}
\centering
\includegraphics[height=7.00cm]{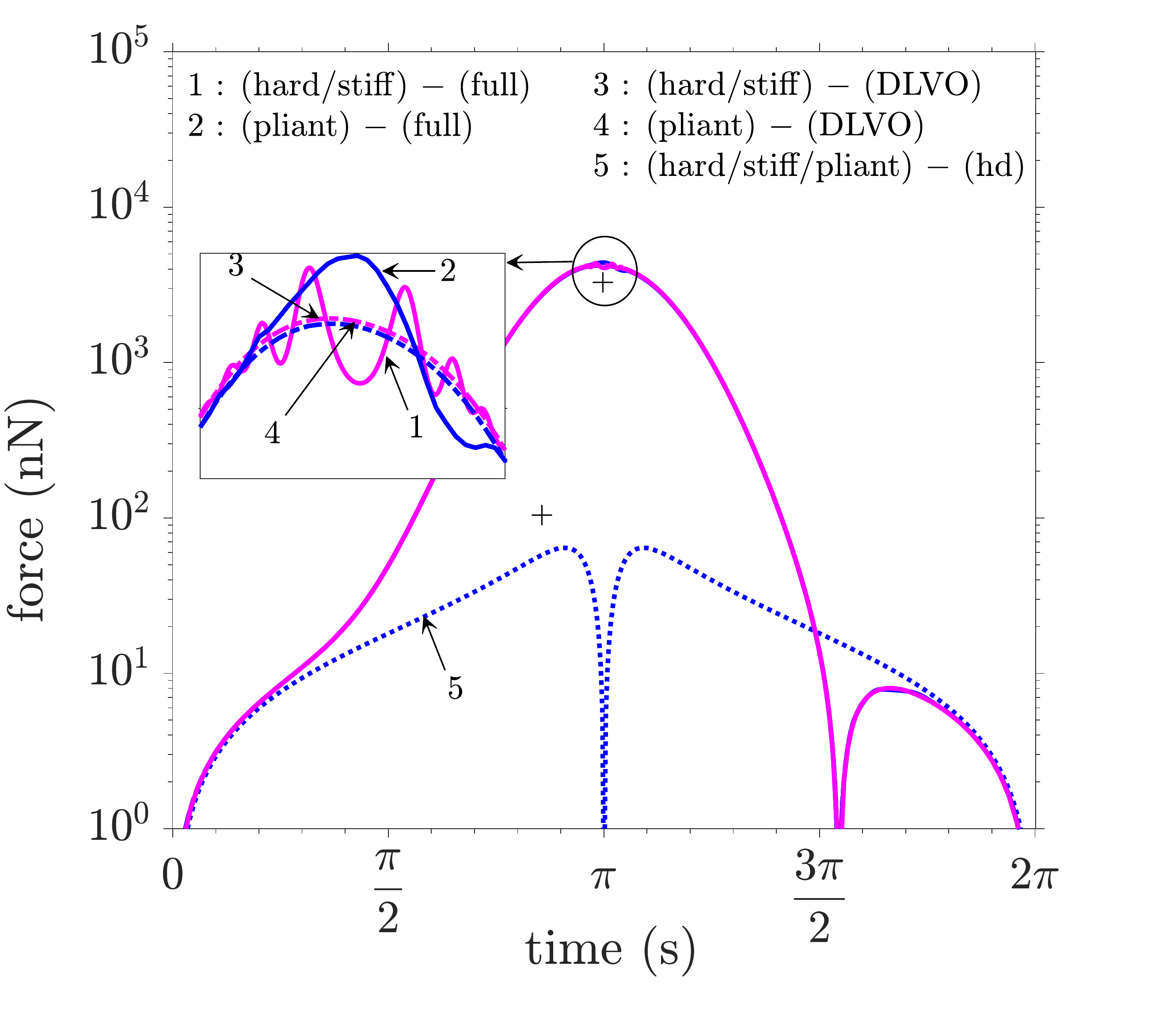}
\caption{\centering}
\label{subfig:F_48d00_stiff_pliant}
\end{subfigure}
\begin{subfigure}[b]{0.5\textwidth}
\centering
\includegraphics[height=7.00cm]{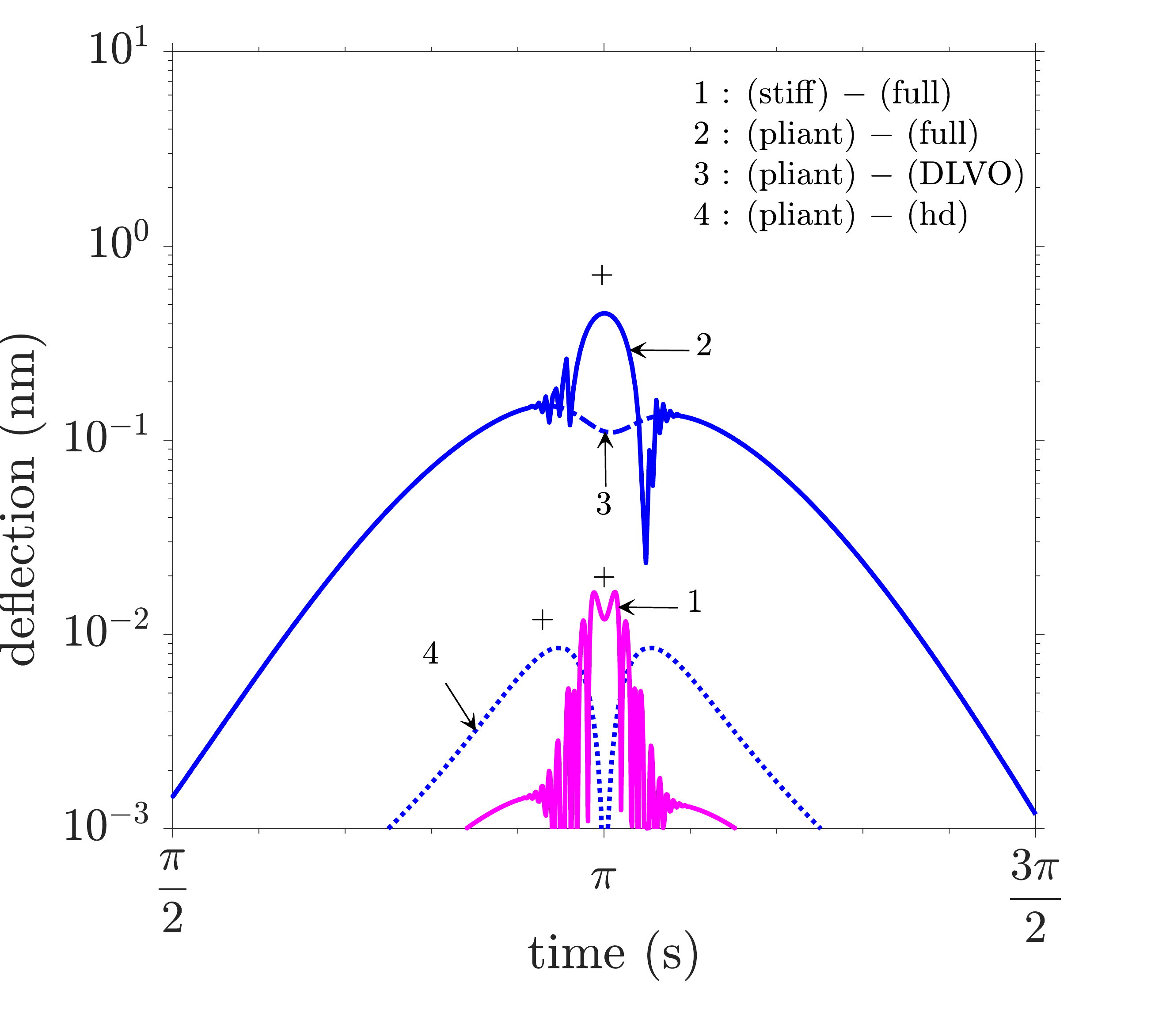}
\caption{\centering}
\label{subfig:lcl_48d00_stiff_pliant}
\end{subfigure}
\caption{Evolution of (a) force between sphere and substrate, and, (b) fluid-substrate inteface deflection at the origin, for the 48 nm amplitude case; solutions for only hard, stiff and pliant substrates are obtained for this amplitude case}
\label{fig:48d00}
\end{figure*}

\begin{figure*}[h!]
\centering
\begin{subfigure}[b]{0.5\textwidth}
\centering
\includegraphics[height=7.00cm]{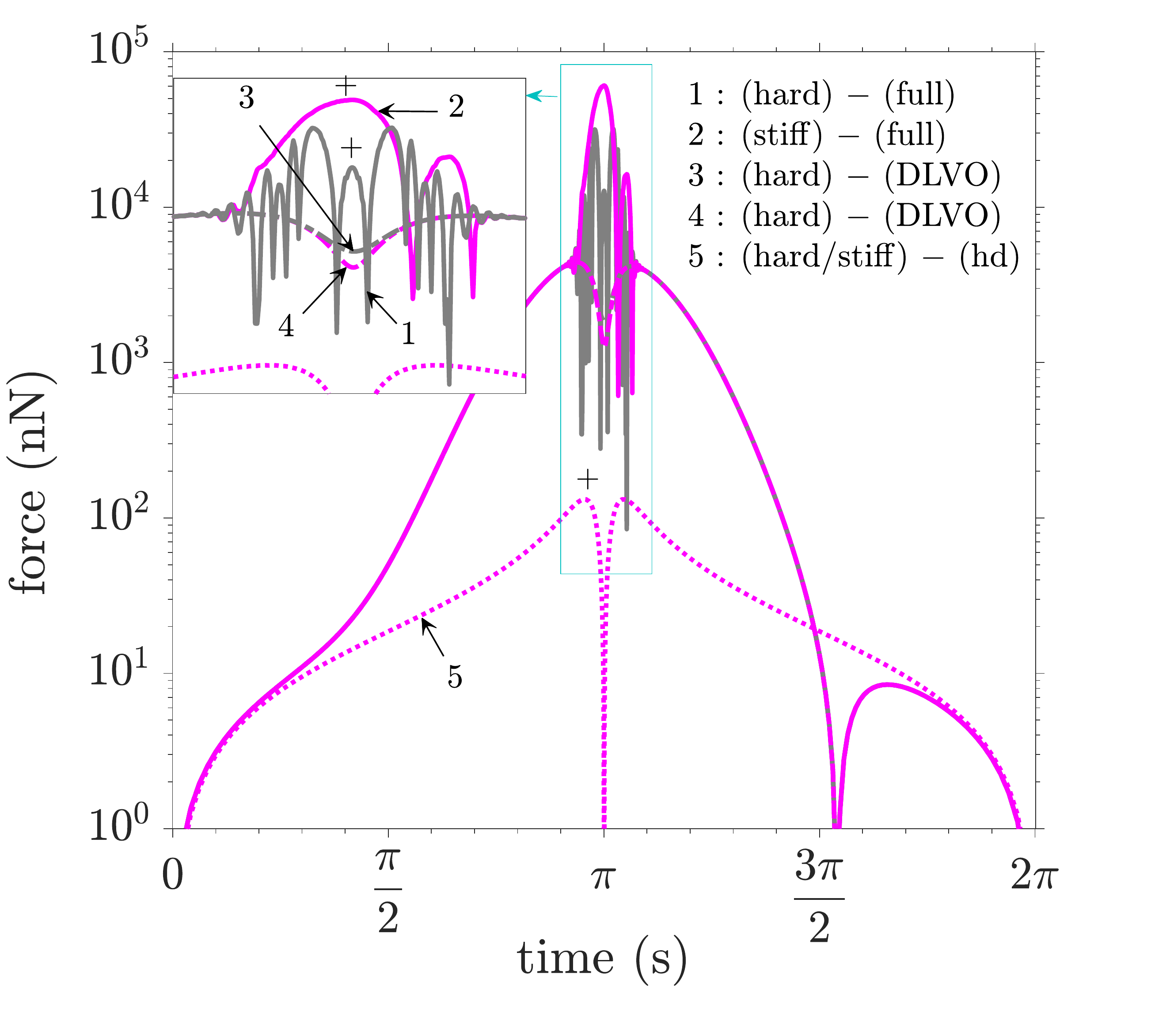}
\caption{\centering}
\label{subfig:F_49d50_rigid_stiff}
\end{subfigure}
\begin{subfigure}[b]{0.5\textwidth}
\centering
\includegraphics[height=7.00cm]{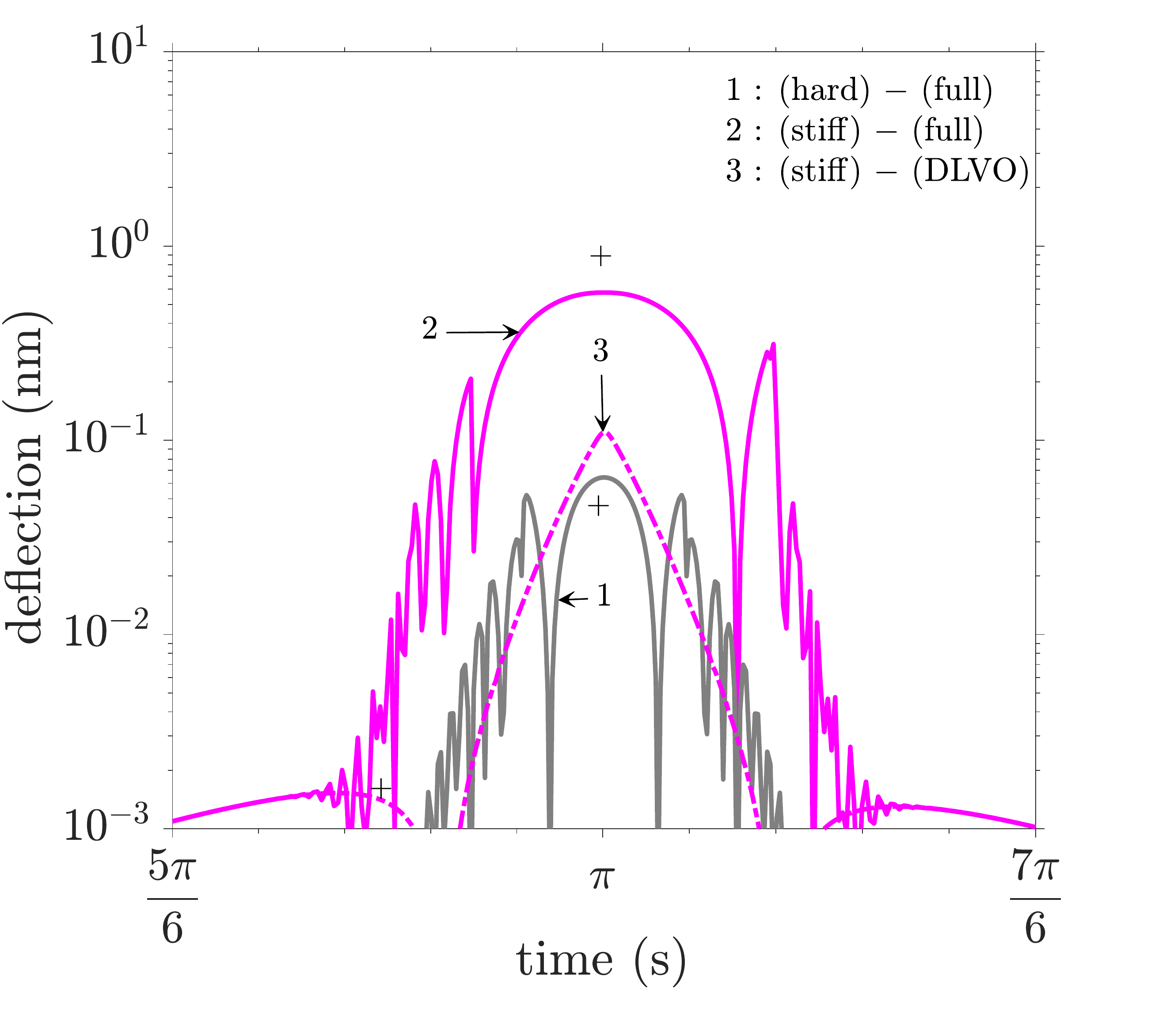}
\caption{\centering}
\label{subfig:lcl_49d50_rigid_stiff}
\end{subfigure}
\caption{Evolution of (a) force between sphere and substrate, and (b) fluid-substrate inteface deflection at the origin, for the 49.5 nm amplitude case; solutions for only hard and stiff substrates are obtained for this amplitude case}
\label{fig:49d50}
\end{figure*}

We investigate the three cases of 40 nm amplitude, 48 nm amplitude and 49.5 nm amplitude. These amplitudes cases are distinct based on which pressure component(s) are significant at and around the origin near mid-oscillation. For the 40 nm amplitude case, it is the EDL disjoining pressure, with the other pressure components negligible; for the 48 nm amplitude case, it is all the three non-hydrodynamic pressure components, with the hydrodynamic pressure negligible; for the 49.5 nm amplitude case, it is the solvation pressure, with the other pressure components negligible. The 40 nm, 48 nm, and 49.5 nm amplitude cases are represented by figures \ref{fig:40d00}, \ref{fig:48d00}, and \ref{fig:49d50} respectively. Each figure has force evolution presented in the left panel and deflection evolution presented in the right panel. All plots are logarithm-scaled on the vertical axis and linear-scaled on the horizontal axis. Absolute values are plotted to accommodate the plots on the log scale. For plots that have negative values, one positive maxima is marked with a `+' sign and every alternate maxima from it is positive. Deflection evolution is shown for only a duration around the mid-oscillation when deflection magnitude is significant. Deflection evolution for substrate$-$solution pairs that have magnitude lesser than 1 pm throughout the oscillation are not presented. A short discussion on evolution of individual pressure components and total pressure at the origin for these three amplitude cases for the full solution of hard substrate is presented in Appendix \ref{sec:ap_press}. We now discuss these amplitude cases one by one.
\subsubsection{EDL Pressure Dominated Amplitude Case}\label{subsubsec:40d00}
This case has $h_0$ = 40 nm, and is represented by figure \ref{fig:40d00}. The dominant pressure component near mid-oscillation is the repulsive EDL disjoining pressure, with the other pressure components being negligible in comparison. The full solution and DLVO solution are identical for each of the substrates. \\
We discuss the force evolution first, presented in figure \ref{subfig:F_40d00_pliant_soft}.\\
Considering the full solution, the force evolution for the hard, stiff and pliant substrates are identical. However, there is contrast between the force evolution for these three and for the soft substrate, particularly near the mid-oscillation. Force evolution for the soft substrate grows to a smaller maxima compared to the other three. This occurs because the fluid-substrate interface for the soft substrate deflects further into the substrate bulk (because of higher flexibility) and hence yields a larger gap-height. This leads to smaller magnitude of the EDL disjoining pressure, and this effect extends along the radial span as well and is pronounced near mid-oscillation. \\
Considering the DLVO solution, the force evolution is identical to that for the full solution for each of the four substrates. This is because the substrate-sphere separation doesn't get small enough for solvation pressure to become significant at any instance during the oscillation. \\
Considering the hd solution, the force evolution for all four substrates are identical. The maximum force for hd solution is about two orders smaller than full solution, indicating the importance of considering non-hydrodynamic force components at such separations. Furthermore, there is significant qualitative contrast between force evolution for the full solution and for the hd solution. While the former is positive for most part of the oscillation and exhibits a significantly higher repulsive maximum than attractive, the latter is anti-symmetric, with the first half-oscillation being repulsive and the second half-oscillation being attractive with equal magnitude. \\
We discuss the deflection evolution next, presented in figure \ref{subfig:lcl_40d00_pliant_soft}. \\
Considering the full solution, the deflection for hard and stiff substrates is negligible throughout the oscillation. The deflection evolution of pliant and soft substrates shows identical qualitative trends. The maxima for the pliant and soft substrates are $\sim$ 80 pm and $\sim$ 5 nm, whose ratio is significantly smaller than the ratio of their $k_{\text{flex}}$. This contrast between the ratio of maxima and the ratio of $k_{\text{flex}}$ is attributable to the larger `push-in' of the soft substrate, the same phenomenon that explained the contrast in force maxima of the first three and the soft substrate. \\
Considering the DLVO solution, as expected, the deflection evolution is identical to that for the full solution for each of the four substrates. \\
Considering the hd solution, the deflection for hard, stiff and pliant substrates is negligible throughout the oscillation. The deflection evolution of soft substrate is qualitatively similar to its force evolution, with maxima being $\sim$ 80 pm. \\
\subsubsection{Intermediate Amplitude Case}\label{subsubsec:48d00}
This case has $h_0$ = 48 nm, and is represented by figure \ref{fig:48d00}. Only the hard, stiff and pliant substrates are studied with this amplitude. Near mid-oscillation, all three non-hydrodynamic pressure components are significant and much larger than hydrodynamic pressure, with solvation pressure being the dominant pressure component at and around the origin.\\
We discuss the force evolution first, presented in figure \ref{subfig:F_48d00_stiff_pliant}. \\
Considering the full solution, the force evolution for the hard and stiff substrates are identical. The force evolution for the pliant substrate is also identical for most part of the oscillation, but near the mid-oscillation, the pliant substrate exhibits lesser fluctuations than the other two. Comparing with the 40 nm amplitude case, while the trends are qualitatively similar, there are two key differences - (i) short-lived but distinct fluctuations appear near mid-oscillation, an indication of solvation pressure becoming significant, and, (ii) the magnitude of maxima is about half an order of magnitude higher. \\
Considering the DLVO solution, the force evolution is identical to that for the full solution for each of the three substrates, except for the lack of fluctuations as can be seen in the inset of figure \ref{subfig:F_48d00_stiff_pliant}. \\
Considering the hd solution, the force evolution for all three substrates are identical. Comparing with the 40 nm amplitude case, while the trends are qualitatively similar, there are two key differences - (i) the maxima are sharper and closer to mid-oscillation, and, (ii) the magnitude of maxima is about half an order higher. These differences are expected, because hydrodynamic pressure follows the expression $\displaystyle \frac{3\sin(t)}{H^2}\cdot\frac{\mu\omega\alpha\epsilon_{0}}{\epsilon^{2}}$, which for a given oscillation amplitude (i.e. for a given $\alpha$) evolves with time according to the expression $\displaystyle \sin(t)\cdot(1+\alpha\cos(t))^{-2}$ and has the pressure-amplitude $\displaystyle \sim 3\mu\omega\alpha\epsilon_{0}$. For a higher $\alpha$, the time-evolution gets sharper and closer to $\displaystyle t=\pi$~s and the pressure-amplitude increases. \\
We discuss the deflection evolution next, presented in figure \ref{subfig:lcl_48d00_stiff_pliant}. \\
Considering the full solution, the deflection for the hard substrate is negligible throughout the oscillation. The deflection evolution of stiff and pliant substrates show identical qualitative trends, with maxima being $\sim$15 pm and $\sim$0.4 nm respectively. The ratio of these maxima is significantly smaller than the ratio of the substrates' $k_{\text{flex}}$, an outcome of the softness-induced `push-in' of the pliant substrate (this observation is in tandem with the deflection evolution contrast of the pliant and soft substrates for the 40 nm amplitude case). Furthermore, as expected, both substrates exhibit fluctuations caused by solvation pressure. There is an interesting qualitative contrast though. The pliant substrate exhibits fluctuation-free smooth deflection evolution for a significant duration near the mid-oscillation, with rapid fluctuations occurring before and after. This observation indicates that the pliant substrate exhibits a `lock-in' with the sphere, meaning that for the said duration near the mid-oscillation, its deflection evolves in such a manner that the sphere-substrate separation remains almost constant. A similar observation is absent for the stiff substrate. \\
Considering the DLVO solution, the deflection evolution is identical to that for the full solution for each of the three substrates, except for the lack of fluctuations.\\
Considering the hd solution, the deflection evolution shows trends similar to force evolution, and is significant only for the pliant substrate and for a small duration near mid-oscillation. Its maximum magnitude is $\sim$ 8 pm. \\
\subsubsection{Solvation Pressure Dominated Amplitude Case}\label{subsubsec:49d50}
This case has $h_0$ = 49.5 nm, and is represented by figure \ref{fig:49d50}. Only the hard and stiff substrates are studied with this amplitude. Near mid-oscillation, and at and around the origin, solvation pressure is much higher than all the other pressure components and is practically equal to the total pressure.\\
We discuss the force evolution first, presented in figure \ref{subfig:F_49d50_rigid_stiff}.\\
Considering the full solution, the force evolution for the hard and stiff substrates are identical for most part of the oscillation. However, near mid-oscillation, the stiff substrate exhibits smaller fluctuations than the hard substrate. This `hard-stiff' contrast is analogous to `stiff-pliant' contrast for the 48 nm amplitude case. Further comparing with the 48 nm amplitude case, there are two key differences - (i) the flutuations in force response near mid-oscillation have much higher magnitude, and (ii) the magnitude of force maxima for full solution is higher by almost two orders of magnitude. Both these differences are attributable to the pronounced effect of solvation pressure, which persists for a higher duration as well as extends farther along the radial span for the 49.5 nm amplitude case in comparison to the 48 nm amplitude case.\\
Considering the DLVO solution, the force evolution is identical to that for the full solution for each of the two substrates, except for the lack of fluctuations as can be seen in the inset of figure \ref{subfig:F_49d50_rigid_stiff}.\\
Considering the hd solution, the force evolution for the hard and stiff substrates is identical. Comparing with the 48 nm amplitude case, while the trends are qualitatively similar, the maxima is significantly sharper and closer to the mid-oscillation. This is an expected observation and is explained based on the expression for hydrodynamic pressure (as discussed in section \ref{subsubsec:48d00} above). The magnitude of maxima is about half an order of magnitude higher than that for the 48 nm amplitude case.\\
We discuss the deflection evolution next, presented in figure \ref{subfig:lcl_49d50_rigid_stiff}.\\
Considering the full solution, the deflection evolution of hard and stiff substrates shows identical qualitative trends, with maxima being $\sim$60 pm and $\sim$0.6 nm respectively. The ratio of the maxima is smaller than the ratio of the substrate $k_{\text{flex}}$, an effect that has been observed for the smaller amplitude cases as well and occurs due to higher push-in of the softer substrate. As expected, both substrates exhibit fluctuations near the mid-oscillation that are caused by solvation pressure. Furthermore, similar to pliant substrate for the 48 nm amplitude case, the stiff substrate exhibits fluctuation-free smooth deflection evolution for a significant duration near the mid-oscillation, with rapid fluctuations before and after.\\
Considering the DLVO solution, the deflection for the stiff substrate has positive values (i.e. into the substrate) of the order of 1 pm for times closer to $\displaystyle t=\frac{5\pi}{6}$ s and $\displaystyle t=\frac{7\pi}{6}$ s. However, there is a large negative maxima at the mid-oscillation, where the `adhesive' deflection grows upto almost 0.1 nm. Deflection for the hard substrate is negligible throughout the oscillation.\\
Considering the hd solution, the deflection for either substrate is negligible throughout the oscillation.\\
\subsubsection{Aggregate Inferences}\label{subsubsec:summ}
While the individual discussion for each amplitude case gives insights into the system behaviour for the respective amplitude, there are a few key aggregate inferences we obtain from examining these evolution trends, discussed below. \\
First, the presence of non-hydrodynamic forces lead to significant magnification in force and deflection response of the system. As can be inferred from figure \ref{fig:49d50}, upto three orders of magnitude amplification occurs in both. Also, for amplitude cases where solvation force is significant, it strongly affects the system behaviour near mid-oscillations, and brings about rapid fluctuations in the force and deflection response. Furthermore, presence of solvation force can be seen to induce deflection in substrates that are effectively rigid otherwise, as can be inferred from deflection evolution for the full solution for hard substrate in figure \ref{subfig:lcl_49d50_rigid_stiff}.\\
Second, for amplitude cases where solvation pressure is significant, the fluctuations in deflection are much higher in magnitude than those in force (compare figure \ref{subfig:F_48d00_stiff_pliant} with figure \ref{subfig:lcl_48d00_stiff_pliant} and figure \ref{subfig:F_49d50_rigid_stiff} with figure \ref{subfig:lcl_49d50_rigid_stiff}). This is attributed to the fact that while the deflection evolution is affected by the pressure at origin, force is affected by the trend of pressure not only at the origin but over the radial span as well. Since the solvation pressure is strong near the origin and reduces to zero with increasing radial distance from the origin (due to increasing sphere-substrate separation), the net effect of considering the radial span is amelioration of the fluctuations. \\
Third, softer substrate exhibit smaller fluctuations in force. They are also prone to a `lock-in' duration near the mid-oscillation (where sphere approach is compensated by substrate deflection and the gap-height remains almost constant with time), which is characterized by smooth fluctuation-free deflection evolution for a significant duration near mid-oscillation. These effects can be observed in the pliant substrate for the 48 nm amplitude case (in figure \ref{subfig:lcl_48d00_stiff_pliant}) and the stiff substrate for the 49.5 nm amplitude case (in figure \ref{subfig:lcl_49d50_rigid_stiff}).\\
Fourth, considering any two substrates, the ratio of deflection for the harder substrate to that for the softer substrate tends to be smaller than the ratio of their $k_{\text{flex}}$ values. This effect is attributed to the fact that the softer substrate exhibits the aforementioned `push-in' effect to a larger extent than the harder substrate. We summarize this `push-in' effect as follows. Approach of the sphere towards the substrate directly contributes to decrease in gap-height. However, this leads to increase in pressure on the substrate, which leads to increase in deflection of the fluid-substrate interface, which contributes to increase in gap-height. In other words, approach of the sphere towards the substrate indirectly contributes to increase in gap-height as well. The aggregate system behaviour is thus a balance between the two opposing effects of the approaching sphere, and the latter effect is stronger for softer substrates. \\
Overall, as one considers increasingly large amplitudes, there are two underlying effects that come into play and whose derivatives are observed in the system behaviour - substrate softness, and, solvation pressure. The effect of substrate softness is summarized as follows. Softness leads to a `push-in' effect (as discussed in third point above), a `lock-in' effect (as discussed in the second point above), and acts to reduce solvation-pressure-induced fluctuations in force response (as discussed in the second point above). Now, the effect of solvation pressure is explained as follows. For the amplitude case of 48 nm, i.e. figure \ref{fig:48d00}, it is evident that when solvation pressure becomes significant, it brings about fluctuations in force as well as deflection response near the mid-oscillation. While these fluctuations are appreciably rapid, their magnitude is not very high, and thus, solvation pressure doesn't lead to any significant amplification in force and deflection response. However, for the higher amplitude case of 49.5 nm, i.e. figure \ref{fig:49d50}, solvation pressure is dominant over other pressure components for a longer duration and over a larger radial span for this amplitude. This leads to fluctuations of higher magnitude as well as substantial amplifications in the force and deflection response. \\
\begin{figure*}[h!]
\centering
\begin{subfigure}[b]{0.5\textwidth}
\centering
\includegraphics[height=7.00cm]{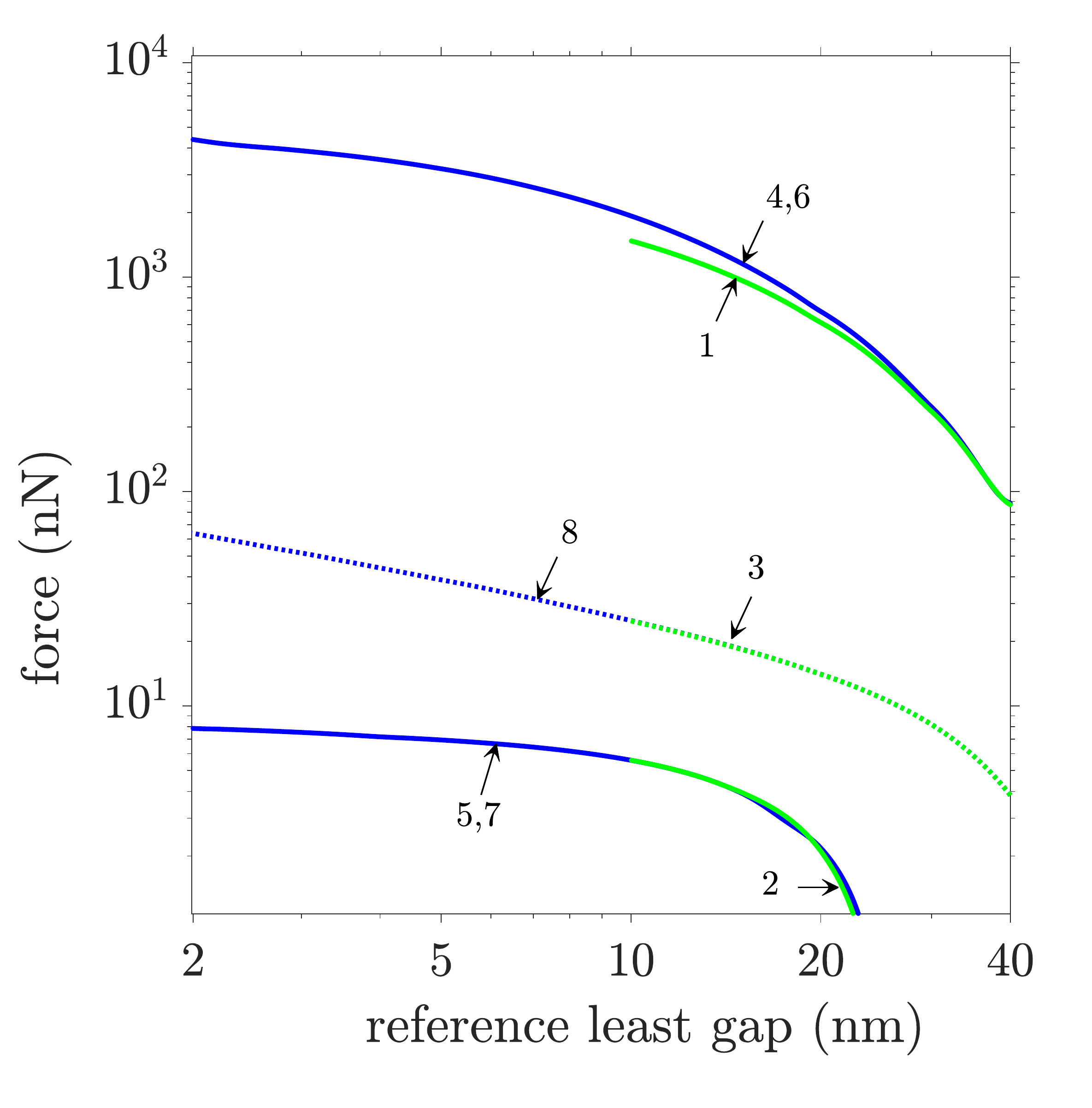}
\caption{\centering}
\label{subfig:F_pliant_soft}
\end{subfigure}
\begin{subfigure}[b]{0.5\textwidth}
\centering
\includegraphics[height=7.00cm]{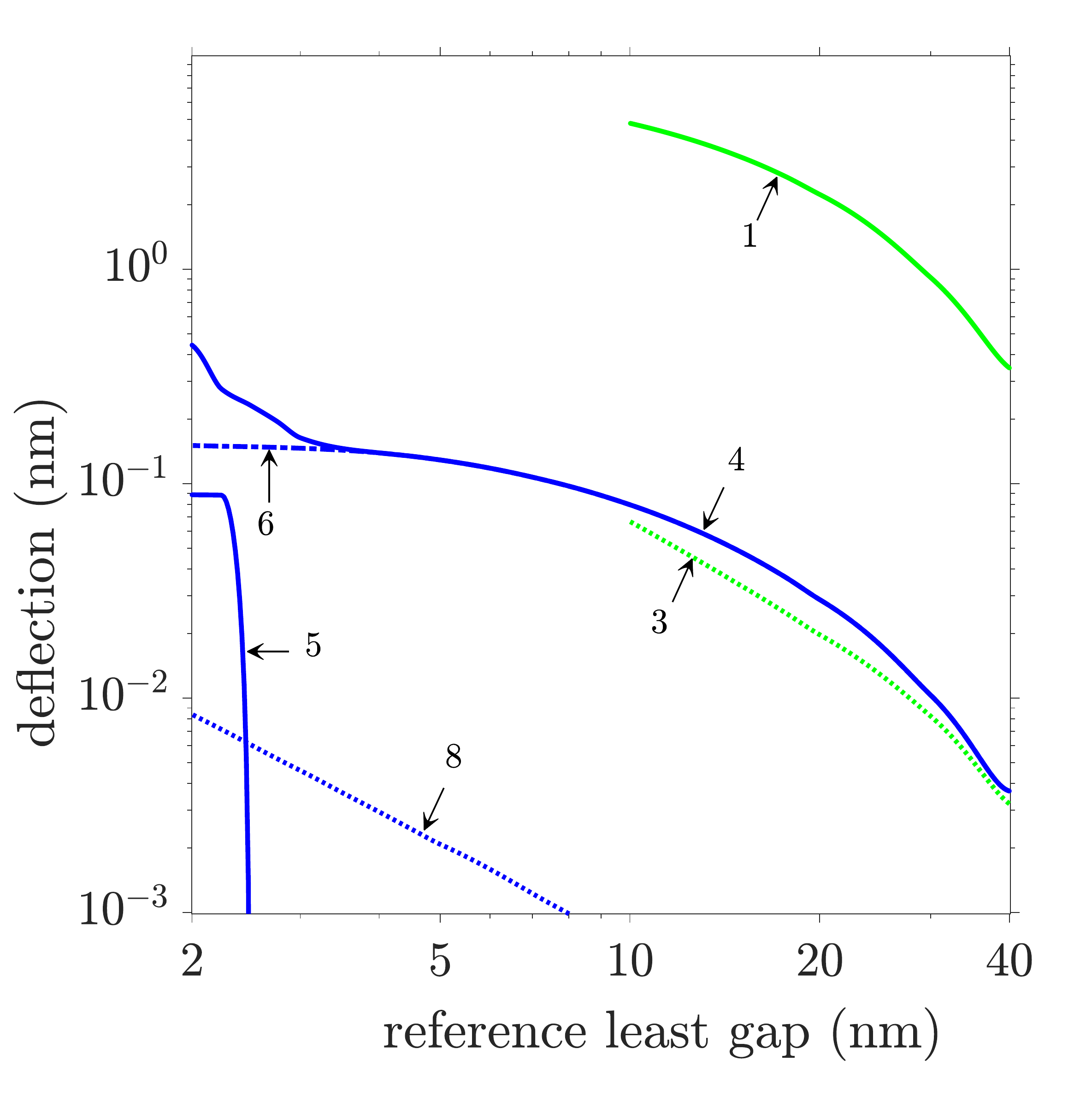}
\caption{\centering}
\label{subfig:l_pliant_soft}
\end{subfigure}
\caption{Variation with reference least gap (the minimum separation of sphere from origin in one complete oscillation, equal to $D-h_{0}$) of maximum repulsive  (labelled with `rep') and maximum attractive (labelled with `att') characteristics over an oscillation of (a) force, and (b) deflection at origin, for soft and pliant substrates; the legend presented on the right is applicable to both the panels. The legend for the figure is as follows: 1 - soft $-$ full $-$ rep, 2 - soft $-$ full $-$ att, 3 - soft $-$ hd, 4 - pliant $-$ full $-$ rep, 5 - pliant $-$ full $-$ att, 6 - pliant $-$ DLVO $-$ rep, 7 - pliant $-$ DLVO $-$ att, 8 - pliant $-$ hd}
\label{fig:aggr_pliant_soft}
\end{figure*}
\subsection{Force and Deflection Characteristics}\label{subsec:characteristics}
To highlight the effects of non-hydrodynamic pressure components, particularly solvation pressure, we examine the variation of maximum repulsive and maximum attractive force and deflection characteristics with oscillation amplitude. These characteristics, for the soft and pliant substrates are presented in figure \ref{fig:aggr_pliant_soft}, and for the stiff and hard substrates in figure \ref{fig:aggr_rigid_stiff}. For either figure, the panel on the left depicts force characteristics and that on the right depicts deflection characteristics. For all the panels, the horizontal axis presents the `reference least gap', the least separation of the sphere from the origin, i.e. $D-h_{0}$, which effectively represents the amplitudes. Both the horizontal and vertical axes in all the panels are logarithm-scaled. In the legends, the label `rep' represents maximum repulsive and the label `att' represents maximum attractive. For hd solution, we emphasize that the maximum repulsive characteristics are exactly equal to the maximum attractive characteristics. Thus hd solution is presented using one plot-line without a `rep' or `att' label. If the deflection characteristics for a substrate$-$solution combination is smaller than 1 pm for all reference least gaps studied, it is not depicted in the figures.\\
\subsubsection{Soft and Pliant Substrates}\label{subsubsec:softpliant}
We first discuss the force and deflection characteristics for the soft and pliant substrates, presented in figure \ref{fig:aggr_pliant_soft}. \\
We consider the soft substrate first. For this substrate, the solvation pressure does not become significant for any reference least gap, as we are restricted to study only amplitudes upto 40 nm (higher amplitude cases require a non-linear constitutive formulation, as formerly specified). Thus, its full solution and DLVO solution are identical.\\
Examining the force characteristics first (plots 1 to 3 in figure \ref{subfig:F_pliant_soft}), both full and hd solutions can be seen to increase monotonically and almost linearly with decreasing reference least gap. Furthermore, the maximum repulsive force for the full solution remain almost two orders of magnitude higher than that for the hd solution. On the other hand, the maximum attractive force for the full solution remain smaller than that for the hd solution, albeit by a lower factor. This is expected because the force over an oscillation is largely influenced by EDL disjoining pressure, and thus predominantly repulsive. There is only a small duration towards the end of the oscillation where the attractive hydrodynamic pressure grows strong enough to counteract the repulsive EDL disjoining pressure and result in a small-magnitude attractive force response (see the the small attractive maxima near $t=2\pi$ s in figure \ref{subfig:F_40d00_pliant_soft}).\\
Examining the deflection characteristics next (plots 1 and 3 in figure \ref{subfig:l_pliant_soft}), they exhibit a monotonic, almost-linear growth with decreasing reference least gap, similar to force. The maximum repulsive deflection for the full solution is almost two orders higher than that for the hd solution. \\
We now consider the pliant substrate. For this substrate, we are able to examine upto 48 nm amplitude case (higher amplitude cases exhibit adhesion-like phenomenon, which is outside the scope of current model, as formerly specified). Thus, solvation pressure becomes significant for $\sim$ 2 to 3 nm reference least gap, and the effects are visible in the deflection characteristics. The trends of force and deflection characteristics, and thus the inferences, are similar to those for the soft substrate. \\
Examining the force characteristics first (plots 4 to 7 in figure \ref{subfig:F_pliant_soft}), the repulsive force for full solution for the pliant substrate are higher in magnitude than those for soft substrate, particularly for reference least gaps closer to 10 nm (compare `pliant $-$ full $-$ rep' and `soft $-$ full $-$ rep' in figure \ref{subfig:l_pliant_soft}). This contrast is attributed to the `pushing-in' of the soft substrate, an effect already discussed in subsection \ref{subsec:amplitude}. \\
Examining the deflection characteristics next (plots 5, 6, and 8 in figure \ref{subfig:l_pliant_soft}), while the qualitative trends for pliant substrate are similar to those for soft substrate, there is an almost two orders of magnitude contrast, an expected observation. For lower than 10 nm amplitude cases, deflection of the pliant substrate retains its monotonic increase. However, maximum repulsive deflection for the full solution exhibits sharp growth between 3 nm and 2 nm reference least gap. This observation is accompanied by the emergence of non-negligible maximum attractive deflection for the full solution. These observations are attributed to the effect of solvation pressure, which becomes significant for reference least gaps smaller than 3 nm. Furthermore, the absence of such trends in force characteristics occurs because deflection responds to any changes in the interplay of pressure components at the origin itself, while force requires these alterations to spread further into the radial span, which would require lower reference least gaps.\\
\subsubsection{Stiff and Hard Substrates}\label{subsubsec:stiffhard}
\begin{figure*}[htb!]
\centering
\begin{subfigure}[b]{0.500\textwidth}
\centering
\includegraphics[height=7.00cm]{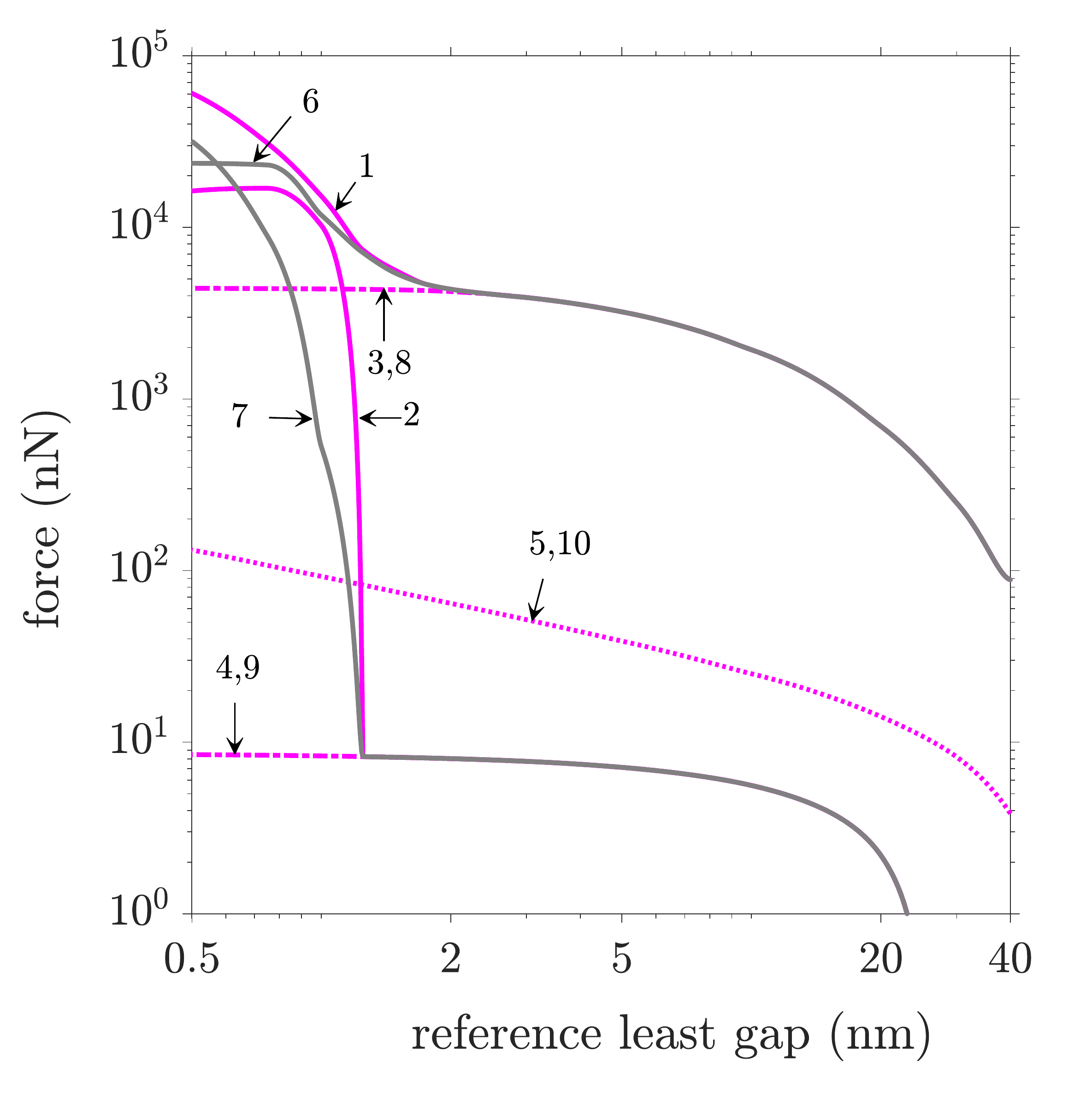}
\caption{\centering}
\label{subfig:F_rigid_stiff}
\end{subfigure}
\begin{subfigure}[b]{0.500\textwidth}
\centering
\includegraphics[height=7.00cm]{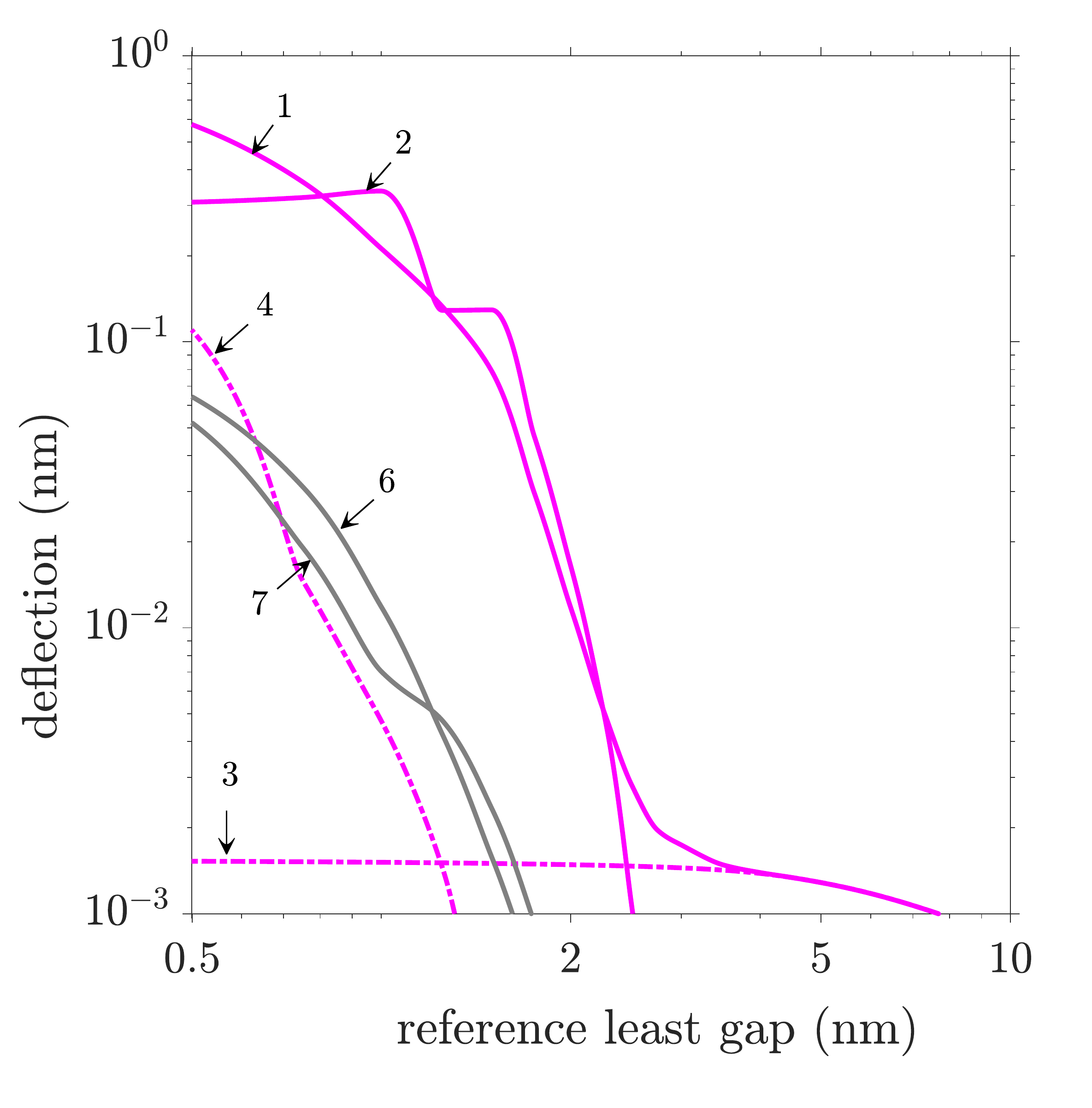}
\caption{\centering}
\label{subfig:l_rigid_stiff}
\end{subfigure}
\caption{Variation with reference least gap (the minimum separation of sphere from origin in one complete oscillation, equal to $D-h_{0}$) of maximum repulsive  (labelled with `rep') and maximum attractive (labelled with `att') characteristics over an oscillation of (a) force, and (b) deflection at origin, for stiff and hard substrates; the legend presented on the right is applicable to both the panels. The legend for the figure is as follows: 1 - stiff $-$ full $-$ rep, 2 - still $-$ full $-$ att, 3 - stiff $-$ DLVO $-$ rep, 4 - still $-$ DLVO $-$ att, 5 - soft $-$ hd, 6 - hard $-$ full $-$ rep, 7 - hard $-$ full $-$ att, 8 - hard $-$ DLVO $-$ rep, 9 - hard $-$ DLVO $-$ att, 10 - hard $-$ hd}
\label{fig:aggr_rigid_stiff}
\end{figure*}

We now discuss the force and deflection characteristics for the stiff and hard substrates, presented in figure \ref{fig:aggr_rigid_stiff}. For both the substrates, we study amplitudes upto 49.5 nm. Hence, we observe the effects of solvation pressure for $\sim$0.5-3 nm reference least gaps, and these effects are visible in both the force and deflection characteristics of each substrate. \\
We consider the stiff substrate first.\\
Examining the force characteristics first (plots 1 to 5 in figure \ref{subfig:F_rigid_stiff}), we observe that both maximum repulsive and maximum attractive force for the full solution show sharp growth in magnitude at low values of reference least gap. The highest values of the maximum repulsive and attractive forces are between 10 $\mu$N and 100 $\mu$N. Two key features of these force characteristics are discussed next. First, comparing DLVO solution with hd solution, it can be clearly seen that the maximum repulsive force for the former stays upto two order of magnitude larger than those for the latter, while the maximum attractive force for the former stays about an order of magnitude smaller than those for the latter. Similar contrast was seen for pliant and soft susbtrates as well, and hence the reasoning presented in the section \ref{subsubsec:softpliant} applies. The persistence of this contrast even for smaller reference least gaps is attributed to the fact that upon approach, while van der Waals pressure does grow higher than EDL disjoining pressure at the origin, the effect does not extend over the radial span strongly enough. The aggregate effect is that force is strongly dominated by EDL disjoining pressure, and hence predominantly repulsive, for all reference least gaps. And second, as we approach smaller reference least gaps (smaller than $\sim$ 1.5 to 2 nm), the amplification in force due to consideration of solvation force (i.e. the contrast between full solution and DLVO solution) is about an order of magnitude for the maximum repulsive and three to four orders of magnitude for the maximum attractive. Furthermore, at 0.5 nm reference least gap, the maximum attractive force and maximum repulsive force are of the same order. This occurs because in contrast to larger reference least gaps that exhibit EDL pressure dominance (implying primarily repulsive force response), reference least gaps approaching 0.5 nm exhibit solvation pressure dominance (implying rapidly oscillating but equal-magnitude repulsive and attractive force response).\\
Examining the deflection characteristics next (plots 1 to 4 in figure \ref{subfig:l_rigid_stiff}), the trends are qualitatively similar to those for force. However, there are a couple of crucial differences. First, the deviation of deflection for the full solution from those for the DLVO solution starts at a higher reference least gap of 3 to 4 nm, compared to deviation for the force which starts at reference least gap of $\sim$ 1.5 to 2 nm. Second, the maximum attractive deflection for DLVO solution grows quite rapidly at reference least gaps smaller than $\sim$ 2 nm, an effect that is absent in force. Both these differences are attributed to the fact that deflection is dependent on the pressure at origin, while force is dependent on the interplay of pressure components along the radial span as well.\\
We now consider the hard substrate.\\
Examining the force characteristics first (plots 6 to 10 in figure \ref{subfig:F_rigid_stiff}), the trends for the hard substrate are qualitatively similar to those for stiff substrate. Similar features are observed and the same explanations hold. The contrast in the force for the full solution for the hard and stiff substrates is attributable to the higher deformability of the latter, which acts to significantly alter the gap-height (particularly for smaller reference least gaps in the range of $\sim$ 0.5 to 1.5 nm) and thus the force response.\\
Examining the deflection characteristics next (plots 6 and 7 in figure \ref{subfig:l_rigid_stiff}), we observe deflection only for the full solution and for reference least gaps lower than 2 nm. Both maximum repulsive and maximum attractive deflection are very similar, and of the same order of magnitude. The trends are similar to that for the stiff substrate, but differing by about an order of magnitude.

\clearpage

\section{Conclusion} \label{sec:concl}

In the current study, the force and deformation characteristics of an ultra-thin soft coating on a rigid platform due to the motion an oscillating rigid sphere and mediated by an intervening fluid have been assessed. There are numerous natural and artificial setups having similar geometry and imposed dynamics at different length scales, and the effects of short-range DLVO and non-DLVO molecular forces become non-negligible for such setups of nanometric length scales. Therefore, a pseudo-continum mathematical model conforming to the traditional soft-lubrication paradigm whilst incorporating the effects of non-hydrodynamic and fluid-structuring forces has been prepared. The solution has been obtained using fundamentals of scaling analysis, and asymptotic and semi-analytical methodologies. Non-hydrodynamic forces are quantified by semi-empirical closed form expressions that appear as additional pressure components in the traction-balance condition at the fluid-substrate interface. \\
Solutions for four substrate materials (hard to soft) are obtained, which are chosen to emphasize the effects of non-hydrodynamic forces as well as substrate softness. Amplitudes ranging from 10 nm to 49.5 nm are studied, i.e. undeformed separations as small as 0.5 nm are studied. The results show that EDL disjoining pressure is dominant at the larger separations and solvation pressure is dominant at the smaller separations, with van der Waals pressure becoming significant at smaller separations but not dominant and hydrodynamic pressure remaining negligible throughout. At very small separations, solvation pressure strongly dominates the system behaviour and appears as virtually the only contributors to the force and deformation characteristics.\\
Solvation pressure is observed to account for about one to two orders of magnitude amplification in repulsive force and about three to four orders of magnitude amplification in attractive force between the surfaces, and, about two to three orders of magnitude amplification in repulsive substrate deformation and about an order of magnitude amplification in attractive substrate deformation at the point of least separation (i.e. at the intersection of sphere's axis with the substrate). We thus bring out giant nonlinear amplifications in the interaction forces between dynamically interacting sphere-and-soft-coating pair separated by a fluid layer spanning over nanometer scales, attributed to a combined consequence of electrostatic, van der Waals and solvation forces predominant over such length scales.\\
A few of the future prospects of our study include poro-elastic modelling for the soft-coating, consideration of non-Newtonian intervening fluid, and consideration of steric nature of electrolytic species. In closing, it is expected that the current work would get assimilated in a developing framework aimed at enabling macroscale modelling towards incorporation of exclusive nanoscale phenomenon like steric effects, hydrophobicity, etc.\cite{Vinogradova2000,Das2011,Chakraborty2013,Naik2017,Karan2018a} in a physically sound yet operationally tractable manner, taking cues from other studies addressing multiscale phenomenon using tools like correction terms and factors \cite{Chakraborty2013c}, order-parameter models \cite{Chakraborty2007,Chakraborty2008,Chakraborty2008b,Chakraborty2012}, coupled mesoscale modelling \cite{Chatterjee2006,Chakraborty2013} and non-classical material models \cite{Kaushik2016,Abhimanyu2016}.\\
\begin{appendices}
\section{Governing Equations}\label{sec:ap_gdes}
The non-dimensionalized governing equations and fluid-substrate interface traction-balance condition are listed below. 
\begin{itemize}
\item Continuity Equation
\begin{equation}
\frac{1}{r} \frac{\partial (ru_{r})}{\partial r} + \frac{\partial u_{z}}{\partial z} = 0
\label{eq:ap_cont}
\end{equation}
\item $r$-Momentum Conservation Equation
\begin{multline}
\epsilon\frac{\rho\omega R^{2}}{\mu}\left[\epsilon\frac{\partial v_{r}}{\partial t} + \alpha\left(v_{r}\frac{\partial v_{r}}{\partial r}+v_{z}\frac{\partial v_{r}}{\partial z} \right)\right] \\ = -\frac{\partial p^{\star}}{\partial r} + \epsilon \left[ \frac{1}{r} \frac{\partial}{\partial r} \left(r \frac{\partial v_{r}}{\partial r}\right) - \frac{v_{r}}{r^{2}} \right] + \frac{\partial^2 v_{r}}{\partial z^{2}}
\label{eq:ap_rmom}
\end{multline}
\item $z$-Momentum Conservation Equation
\begin{multline}
\epsilon^{2}\frac{\rho\omega R^{2}}{\mu}\left[\epsilon\frac{\partial v_{z}}{\partial t} + \alpha\left(v_{r}\frac{\partial v_{z}}{\partial r}+v_{z}\frac{\partial v_{z}}{\partial z} \right)\right] = \\ -\frac{\partial p^{\star}}{\partial z} + \epsilon^{2} \left[ \frac{1}{r} \frac{\partial}{\partial r} \left(r \frac{\partial v_{z}}{\partial r}\right) \right] + \epsilon\frac{\partial^2 v_{r}}{\partial z^{2}}
\label{eq:ap_zmom}
\end{multline}
\item Mechanical Equilibrium Equation ($r$-component)
\begin{multline}
(1-2\nu)\frac{\partial^2 u_{r}}{\partial y^2} + \gamma \left[\frac{\partial^2 u_{y}}{\partial r \partial y}\right] + \\ \gamma^{2}\left[2(1-2\nu)(1-\nu) \left(\frac{\partial^2 u_{r}}{\partial r^{2}} + \frac{1}{r} \frac{\partial u_{r}}{\partial r} - \frac{u_{r}}{r} \right)\right] = 0
\label{eq:ap_rdisp}
\end{multline}
\item Mechanical Equilibrium Equation ($y$-component)
\begin{multline}
\frac{\partial^2 u_{y}}{\partial y^2} + \gamma \left[\frac{1}{2(1-\nu)}\left(\frac{\partial^2 u_{r}}{\partial r \partial y} + \frac{\partial u_{r}}{\partial y}\right)\right] + \\ \gamma^{2} \left[ \frac{(1-2\nu)}{2(1-\nu)} \left(\frac{\partial^2 u_{y}}{\partial r^{2}} + \frac{1}{r} \frac{\partial u_{y}}{\partial r} \right)\right] = 0
\label{eq:ap_ydisp}
\end{multline}
\item Fluid-Substrate Interface Traction-Balance Condition ($r$-component)
\begin{multline}
(1-2\nu)\frac{\partial u_{r}}{\partial y} + \gamma \left[(1-2\nu)\frac{\partial u_{y}}{\partial r} - 2 \kappa \nu \frac{\partial u_{y}}{\partial y} \frac{\partial u_{y}}{\partial r} \right] - \\ \gamma^{2}\kappa \left[ 2(1-\nu) \frac{\partial u_{r}}{\partial r}\frac{\partial u_{y}}{\partial r} + 2\nu \frac{u_{r}}{r} \frac{\partial u_{y}}{\partial r} \right] = \\ \frac{2\mu\omega\alpha\epsilon_{0}(1+\nu)(1-2\nu)}{\epsilon^{2}\kappa E_{y}} \cdot \\ \left[\gamma\kappa \left\{ p\frac{\partial u_{y}}{\partial r}-2\epsilon \frac{\partial v_{r}}{\partial r}\frac{\partial u_{y}}{\partial r} \right\} + \epsilon^{\frac{1}{2}}\frac{\partial v_{r}}{\partial z} + \epsilon^{\frac{3}{2}}\frac{\partial v_{z}}{\partial r} \right]
\label{eq:ap_rtrac}
\end{multline}
\item Fluid-Substrate Interface Traction-Balance Condition ($y$-component)
\begin{multline}
\frac{\partial u_{y}}{\partial y} + \gamma \left[\frac{\nu}{1-\nu}\left(\frac{\partial u_{r}}{\partial r}+\frac{u_{r}}{r}\right) - \kappa \frac{1-2\nu}{2(1-\nu)} \frac{\partial u_{r}}{\partial y} \frac{\partial u_{y}}{\partial r} \right] - \\ \gamma^{2}\kappa \frac{(1-2\nu)}{2(1-\nu)} \left(\frac{\partial u_{y}}{\partial r}\right)^{2} = \\ -\frac{\mu\omega\alpha\epsilon_{0}}{\epsilon^{2}\kappa E_{y}}\frac{(1+\nu)(1-2\nu)}{(1-\nu)} \cdot \\ \left[p+\gamma\kappa \left\{ \epsilon^{\frac{1}{2}}\frac{\partial v_{r}}{\partial z}\frac{\partial u_{y}}{\partial r} + \epsilon^{\frac{3}{2}}\frac{\partial v_{z}}{\partial r}\frac{\partial u_{y}}{\partial r} \right\} - 2\epsilon\frac{\partial v_{z}}{\partial z} \right]
\label{eq:ap_ytrac}
\end{multline}
\end{itemize}
The other boundary conditions are,
\begin{itemize}
\item no-slip and no-penetration conditions at fluid-sphere interface and fluid-substrate interface
\item zeta-potential at fluid-sphere interface and fluid-substrate interface
\item electroneutral number-density for each electrolytic species at the radial far-end
\item zero-displacement condition at the substrate-platform interface
\item deformation and velocity fields and hydrodynamic pressure reduced to zero at radial far-end
\item zero $r$-deformation and $r$-velocity and zero radial-derivative of $y$-deformation and $z$-velocity at centerline ($r$=0)
\end{itemize}
\begin{table}[h!]
\small
\caption{Perturbation split of Reynolds equation, its boundary conditions, non-hydrodynamic pressure components, and expression for deflection about the small parameter $\displaystyle \eta$}
\label{tab:ap_perthd}
\begin{tabular*}{1\textwidth}{@{\extracolsep{\fill}}ll}
\\[5pt]
\hline
\\[1pt]
\textbf{Order}  		&	\textbf{Split} 																																	\\[5pt]
\hline
~						& 	~ 																																				\\
$\displaystyle \eta^0$	& 	$\displaystyle -\sin(t) = \frac{1}{12r} \frac{\partial}{\partial r} \left[ rH^3\frac{\partial p^{\star}_{(0)}}{\partial r}\right]$; 			\\[15pt]
& 	$\displaystyle \frac{\partial p^{\star}_{(0)}}{\partial r} = 0$ at $r=0$, $\displaystyle p^{\star}_{(0)} = 0$ at $r=1$.							\\[15pt]
&	$\displaystyle \pi_{\text{DL}(0)} = \frac{2 \bar{\epsilon}\bar{\epsilon}_{0}\epsilon^2K^2\zeta^2}{\mu\omega\alpha\epsilon_{0}} \ \exp\left(-\epsilon K R H\right)$																																			\\[15pt]
&	$\displaystyle \pi_{\text{vdW}(0)} = -\frac{A_{\text{sfw}}}{6\pi\epsilon\epsilon_{0}\alpha\mu\omega R^3} \frac{1}{H^3}$\\[15pt]
&	$\displaystyle \pi_{\text{S}(0)}=-\frac{\epsilon^{2}\Lambda}{\mu\omega\alpha\epsilon_{0}}\exp\left(-\frac{\epsilon RH}{s}\right)\cos\left(\frac{2\pi\epsilon RH}{s}+\phi\right)$																																					\\[15pt]
& 	$\displaystyle l_{(0)} = \frac{\mu\omega\alpha\epsilon_{0}}{\epsilon^2\kappa E_{y}}\frac{(1+\nu)(1-2\nu)}{(1-\nu)}p_{(0)} = p_{(0)}$ 						\\[15pt]
$\displaystyle \eta^1$	&	$\displaystyle \frac{\epsilon}{\alpha\epsilon_{0}}\frac{\partial l_{(0)}}{\partial t}=\frac{1}{12r}\frac{\partial}{\partial r}\left[rH^3 
\frac{\partial p^{\star}_{(1)}}{\partial r} \right]+\frac{1}{4r}\frac{\partial}{\partial r} \left[rH^2l_{(0)}\frac{\partial p^{\star}_{(0)}}{\partial r}\right]$; 																																	\\[15pt]
&	$\displaystyle \frac{\partial p^{\star}_{(1)}}{\partial r} = 0$ at $r=0$, $\displaystyle p^{\star}_{(1)} = 0$ at $r=1$.							\\[15pt]
&	$\displaystyle \pi_{\text{DL}(1)}=-\epsilon 		KRl_{(0)}\pi_{\text{DL}(0)}$ 																														\\[15pt]
&	$\displaystyle \pi_{\text{vdW}(1)}=-\frac{3l_{(0)}}{H} \pi_{\text{vdW}(0)}$ \\[15pt]
&	$\displaystyle \pi_{\text{S}(1)} = \frac{\epsilon\Lambda Rl_{(0)}}{s} \left[1+2\pi\tan\left(\frac{2\pi \epsilon RH}{s}+\phi \right)\right]\pi_{\text{S}(0)}$ 												\\[15pt]
& $\displaystyle l_{(1)} = \frac{\mu\omega\alpha\epsilon_{0}}{\epsilon^2\kappa E_{y}}\frac{(1+\nu)(1-2\nu)}{(1-\nu)}p_{(1)} = p_{(1)}$ 						\\[15pt]
\hline
\end{tabular*}
\end{table}
The equations for these boundary-conditions are straightforward and hence not presented here. The deformation governing equations, equations \eqref{eq:ap_rdisp} and \eqref{eq:ap_ydisp}, are the non-dimensionalized version of the two components of the mechanical equilibrium euqation, $\nabla'\cdot\underline{\underline{\sigma}}' = 0$. Similarly, equations \eqref{eq:ap_rtrac} and \eqref{eq:ap_ytrac} are the components of $\displaystyle \underline{\underline{\sigma}}'\cdot \hat{n}' = \underline{\underline{\sigma}}_{F}'\cdot\hat{n}'$ that are subsequently subjected to non-dimensionalization, where the substript $F$ signifies the fluid-domain counterpart, and $\displaystyle \hat{n}' = -\hat{e}_z+\frac{\partial u_y'}{\partial r'}\hat{e}_r$ is the normal vector to the fluid-substrate interface. The stress $\underline{\underline{\sigma}}'$ in terms of displacement field $\vec{u}'$ for a linear-elastic substrate is given as,
\begin{equation}
\underline{\underline{\sigma}}' = \frac{\nu E_y}{(1+\nu)(1-2\nu)}\nabla'\cdot\vec{u}' \underline{\underline{I}} + \frac{E_y}{2(1+\nu)}\left(\nabla'\vec{u}' + (\nabla'\vec{u}')^T\right),
\label{eq:ap_linel_SS}
\end{equation}
where, $E_y$ is Young's modulus and $\nu$ is Poisson's ratio. The primes denote that the terms are in their dimensional form.\\
Keeping only terms that are unity-ordered, equations \eqref{eq:ap_cont} to \eqref{eq:ap_ytrac} get simplified to,
\begin{itemize}
\item Continuity Equation (simplified)
\begin{equation}
\frac{1}{r} \frac{\partial (ru_{r})}{\partial r} + \frac{\partial u_{z}}{\partial z} = 0
\label{eq:ap_cont_simp}
\end{equation}
\item $r$-Momentum Conservation Equation (simplified)
\begin{equation}
0 = -\frac{\partial p^{\star}}{\partial r} + \frac{\partial^2 v_{r}}{\partial z^{2}}
\label{eq:ap_rmom_simp}
\end{equation}
\item $z$-Momentum Conservation Equation (simplified)
\begin{equation}
0 = \frac{\partial p^{\star}}{\partial z}
\label{eq:ap_zmom_simp}
\end{equation}
\item Mechanical Equilibrium Equation ($r$-component, simplified)
\begin{equation}
\frac{\partial^2 u_{r}}{\partial y^2} = 0
\label{eq:ap_rdisp_simp}
\end{equation}
\item Mechanical Equilibrium Equation ($y$-component, simplified)
\begin{equation}
\frac{\partial^2 u_{y}}{\partial y^2} = 0
\label{eq:ap_ydisp_simp}
\end{equation}
\item Fluid-Substrate Interface Traction-Balance Condition ($r$-component, simplified)
\begin{equation}
\frac{\partial u_{r}}{\partial y} = 0
\label{eq:ap_rtrac_simp}
\end{equation}
\item  Fluid-Substrate Interface Traction-Balance Condition ($y$-component, simplified)
\begin{equation}
\frac{\partial u_{y}}{\partial y} = -\frac{\mu\omega\alpha\epsilon_{0}}{\epsilon^{2}\kappa E_{y}}\frac{(1+\nu)(1-2\nu)}{(1-\nu)}p
\label{eq:ap_ytrac_simp}
\end{equation}
\end{itemize}
From these simplified equations, the Reynolds equation and its boundary conditions, equations \eqref{eq:Reeq}, \eqref{eq:pbc} and \eqref{eq:pbc1} emerge as the representative of the flow dynamics in the problem. The total pressure, given in equation \eqref{eq:ptot}, is the combination of hydrodynamic pressure and three non-hydrodynamic pressure components, with the three non-hydrodynamic pressure components given by expressions \eqref{eq:piDL}, \eqref{eq:pivdW}, and \eqref{eq:piS}. The solution for $u_r$ is obtained as zero, and the solution for $u_y$ (and resultantly the deflection $l$) is obtained as given in equation \eqref{eq:uy} (and equation \eqref{eq:l}). The expressions for $p_i$'s (total pressure, hydrodynamic pressure and non-hydrodynamic pressure components) and $l$ are subjected to a perturbation approximation in the parameter $\eta$ (as shown in equations \eqref{eq:pertAp} and \eqref{eq:pertAl}) for the methodology presented in subsection \ref{subsec:classpert}. The perturbation splits of Reynolds equation, its boundary conditions, non-hydrodynamic pressure components, and expression for deflection are presented in table \ref{tab:ap_perthd}.
\section{Incompressible Substrate}\label{sec:ap_incompr}
A caveat in the set of equations in appendix \ref{sec:ap_gdes}, and thus in the mathematical forumation of this article, is that the substrate deformation equations and boundary conditions (i.e. equations \eqref{eq:ap_rdisp} to \eqref{eq:ap_ytrac}) are simplified (to equations \eqref{eq:ap_rdisp_simp} to \eqref{eq:ap_ytrac_simp}) assuming (a) equal scale for $r$-deformation and $y$-deformation during non-dimensionalization, and (b) $(1-2\nu) \not\ll 1$. While the former is a characteristic of the system response, the latter is a material property and therefore a given parameter.\\
We focus on assumption (a), and to probe it further, we assume that  O($u_r)$ = $\Gamma $ O($u_y$). With this operation, and without commenting on the magnitude of $\Gamma$, equation \eqref{eq:ap_rdisp_simp} to \eqref{eq:ap_ytrac_simp} get transformed to,
\begin{itemize}
\item Mechanical Equilibrium Equation ($r$-component, simplified)
\begin{equation}
(1-2\nu)\Gamma\frac{\partial^2 u_{r}}{\partial y^2} + \gamma \frac{\partial^2 u_{y}}{\partial r \partial y} = 0
\label{eq:ap_rdisp_simp1}
\end{equation}
\item $y$-Deformation Equation ($y$-component, simplified)
\begin{equation}
\frac{\partial^2 u_{y}}{\partial y^2} + \gamma \Gamma \left[\frac{1}{2(1-\nu)}\left(\frac{\partial^2 u_{r}}{\partial r \partial y} + \frac{\partial u_{r}}{\partial y}\right)\right] = 0
\label{eq:ap_ydisp_simp1}
\end{equation}
\item Fluid-Substrate Interface Traction-Balance Condition ($r$-component, simplified)
\begin{equation}
\Gamma\frac{\partial u_{r}}{\partial y} + \gamma \frac{\partial u_{y}}{\partial r} =  2\frac{\mu\omega\alpha\epsilon_{0}}{\epsilon^{2}\kappa E_{y}}(1-\nu)\gamma\kappa  p\frac{\partial u_{y}}{\partial r}
\label{eq:ap_rtrac_simp1}
\end{equation}
\item Fluid-Substrate Interface Traction-Balance Condition ($y$-component, simplified)
\begin{equation}
\frac{\partial u_{y}}{\partial y} + \gamma\Gamma \frac{\nu}{(1-\nu)}\left(\frac{\partial u_{r}}{\partial r}+\frac{u_{r}}{r}\right) =  -\frac{\mu\omega\alpha\epsilon_{0}}{\epsilon^{2}\kappa E_{y}}\frac{(1+\nu)(1-2\nu)}{(1-\nu)}p
\label{eq:ap_ytrac_simp1}
\end{equation}
\end{itemize}
Focussing on the deflection, i.e. the $y$-displacement at the interface, we see that the simplified $y$-components of mechanical equilibrium equation and fluid-substrate interface traction-balance condition (equations \eqref{eq:ap_ydisp_simp1} and \eqref{eq:ap_ytrac_simp1}) stay de-coupled from the solution for $r$-displacement field so long as $\displaystyle \Gamma\ll\frac{1}{\gamma}$. Thus, the solution for deflection in section \ref{sec:soln} and the results in \ref{sec:results} continue to be valid so long as $\displaystyle \Gamma \ll \frac{1}{\gamma}$. In contrast, when $\displaystyle \Gamma \sim \frac{1}{\gamma}$, equations \eqref{eq:ap_rdisp_simp1} to \eqref{eq:ap_ytrac_simp1} are strongly coupled and not reducible to a de-coupled form. Such a coupled set of equations for the axisymmetric system considered here can be solved employing a Hankel-transformation approach \cite{Harding1945,Li1997,Leroy2011,Wang2017a}, which is a scope for further generalization of current study. \\
However, we attempt a scaling analysis for such a system, with the intent of drawing insights regarding applicability of current formulation. Keeping in view that $\Gamma$ represents the scale for $r$-displacement, its scale should come out of the governing equation for the same. Thus, examining equation \eqref{eq:ap_rdisp_simp1}, it is posited that $r$-displacement comes out of the interaction of the two terms therein, and hence scaling the two terms equally gives,
\begin{equation}
\Gamma = \frac{\gamma}{(1-2\nu)}.
\label{eq:ap_rdisp_scale_1}
\end{equation}
However, as one considers a substrate that is sufficiently close to incompressible, the governing equation for $r$-displacement becomes the condition for incompressibility (primes signifying that the terms are dimensional), 
\begin{equation}
\frac{\partial u_r'}{\partial r'}+\frac{u_r'}{r'}+\frac{\partial u_y'}{\partial y'}=0,
\label{eq:ap_incompr_dim}
\end{equation}
the scale for $\Gamma$ comes from equal scaling of its two terms upon non-dimensionalization, giving,
\begin{equation}
\Gamma = \frac{1}{\gamma},
\label{eq:ap_rdisp_scale_2}
\end{equation}
and the terms in equation \eqref{eq:ap_rdisp_simp1} would be left to play out `spontaneously'. In summary, as one considers values of $\nu$ approaching 0.5 (i.e. substrate behaviour approaching incompressibility), $\Gamma$, following the equation \eqref{eq:ap_rdisp_scale_1}, is initially $\ll 1$, then $\sim 1$, and then  $\gg 1$ until $\displaystyle \frac{\gamma}{(1-2\nu)}$ (RHS of equation \eqref{eq:ap_rdisp_scale_1}) has grown to be equal to $\displaystyle \frac{1}{\gamma}$. As one continues to takes $\nu$ even closer to 0.5, $\Gamma$ is given by equation \eqref{eq:ap_rdisp_scale_2} and is evidently independent of the substrate material properties, substrate deformation characteristics now exhibiting an `incompressible substrate limit'. Furthermore, considering equation \eqref{eq:ap_rdisp_scale_1}, it can be deduced that $\displaystyle \Gamma < 0.1\cdot \frac{1}{\gamma}$ (which has been shown to be the condition for the formulation and solution methodology employed in current study to hold applicable) is equivalent to,
\begin{equation}
\nu < \frac{1-10\gamma^2}{2}. 
\label{eq:ap_applcond}
\end{equation}
It is emphasized that since we are employing a time-dependent scaling, the restriction in equation \eqref{eq:ap_applcond} needs to be applied as per the largest value of $\gamma$ in an oscillation. For the values of substrate thickness, sphere radius and mean sphere-origin separation taken in table \ref{tab:paramtaken}, the maximum allowed value of $\nu$ is 0.468, 0.493, and 0.499 for 49.50 nm, 48.25 nm, and 37.5 nm amplitude oscillations respectively.\\
To assess the behaviour of a perfectly incompressible substrate for our physical setup, we present a scaling analysis employing an alternate constitutive formulation comprising an arbitrary solid pressure that is applicable for incompressible substrates \citep{Ferry1980,Mahadevan2005,Batra2006,Goriely2006,Bijelonja2006,Rallabandi2017,Pandey2016}, given as,
\begin{equation}
\underline{\underline{\sigma}}' = -p_S' \underline{\underline{I}} + \frac{E_y}{3}\left(\nabla'\vec{u}' + (\nabla'\vec{u}')^T\right),
\label{eq:ap_incompr_SS}
\end{equation}
where $p_S'$ is the solid pressure. This solid pressure becomes an additional unknown that has be solved for, and the incompressiblity condition, equation \eqref{eq:ap_incompr_dim} becomes the additional equation required. Equation \eqref{eq:ap_incompr_dim} will give the scale of $\Gamma$ as presented in equation \eqref{eq:ap_rdisp_scale_2}. We next non-dimensionalize the components of mechanical equilibrium equation and fluid-substrate interface traction-balance condition as done in appendix \ref{sec:ap_gdes}, but employing the constitutive relation \eqref{eq:ap_incompr_SS}, considering $\displaystyle \Gamma\lambda(t)$ as the scale for $u_r$ rather than $\lambda(t)$, and considering $p_c$ as the scale for the solid pressure $p_S$. The obtained equations are,
\begin{itemize}
\item Mechanical Equilibrium Equation ($r$-component)
\begin{multline}
-\frac{3\gamma^2 p_c}{\kappa E_y}\frac{\partial p_S}{\partial r} + \frac{\partial^2 u_{r}}{\partial y^2} + \\ \gamma^2 \left[2\frac{\partial^2 u_{r}}{\partial r^{2}} + \frac{\partial^2 u_{y}}{\partial r \partial y} + \frac{2}{r}\left(\frac{\partial u_{r}}{\partial r} - \frac{u_{r}}{r}\right)\right]  = 0
\label{eq:ap_rdisp_incompr}
\end{multline}
\item Mechanical Equilibrium Equation ($y$-component)
\begin{multline}
-\frac{3}{2}\frac{p_c}{\kappa E_y}\frac{\partial p_S}{\partial y} +  \left[2\frac{\partial^2 u_{y}}{\partial y^2} + \frac{\partial^2 u_r}{\partial r \partial y} + \frac{1}{r}\frac{\partial u_r}{\partial y} \right] + \\ \gamma^2\left[ \frac{\partial^2 u_y}{\partial r^2} + \frac{1}{r}\frac{\partial u_r}{\partial y} \right] = 0
\label{eq:ap_ydisp_incompr}
\end{multline}
\item Fluid-Substrate Interface Traction-Balance Condition ($r$-component)
\begin{multline}
\frac{\partial u_r}{\partial y} + \gamma^2 \frac{\partial u_y}{\partial r} + \kappa p_S \frac{\partial u_y}{\partial r} - 2\gamma^2\kappa \frac{\partial u_r}{\partial r} \frac{\partial u_y}{\partial r} =  \frac{3\mu\omega\alpha\epsilon_{0}\gamma}{\epsilon^{2}\kappa E_{y}}\cdot \\ \left[\gamma\kappa \left\{ p\frac{\partial u_{y}}{\partial r}-2\epsilon \frac{\partial v_{r}}{\partial r}\frac{\partial u_{y}}{\partial r} \right\} + \epsilon^{\frac{1}{2}}\frac{\partial v_{r}}{\partial z} + \epsilon^{\frac{3}{2}}\frac{\partial v_{z}}{\partial r} \right]
\label{eq:ap_rtrac_incompr}
\end{multline}
\item Fluid-Substrate Interface Traction-Balance Condition ($y$-component)
\begin{multline}
p_S + \gamma \frac{\partial u_r}{\partial y} - 2\gamma^2 \frac{\partial u_y}{\partial y}+ \gamma^3 \frac{\partial u_y}{\partial r} =  \frac{3\mu\omega\alpha\epsilon_{0}\gamma^2}{\epsilon^2\kappa E_y} \cdot \\ \left[p+\gamma\kappa \left\{ \epsilon^{\frac{1}{2}}\frac{\partial v_{r}}{\partial z}\frac{\partial u_{y}}{\partial r} + \epsilon^{\frac{3}{2}}\frac{\partial v_{z}}{\partial r}\frac{\partial u_{y}}{\partial r} \right\} - 2\epsilon\frac{\partial v_{z}}{\partial z} \right]
\label{eq:ap_ytrac_incompr}
\end{multline}
An additional equation appears, i.e. the non-dimensionalized form of equation \eqref{eq:ap_incompr_dim}, that is applicable for the substrate domain.
\item Incompressibility Condition
\begin{equation}
\frac{\partial u_{r}}{\partial u_{r}} + \frac{u_{r}}{r} + \frac{\partial u_{y}}{\partial y} = 0
\label{eq:ap_incompr_incompr}
\end{equation}
\end{itemize}
The set of equations \eqref{eq:ap_rdisp_incompr} to \eqref{eq:ap_incompr_incompr} are reminiscent of the non-dimensionalized continuity and Stokes equation for lubrication flows. Therefore, taking cue from the same, we have,
\begin{equation}
\frac{3\gamma^2p_c}{\kappa E_y} = 1 \implies p_c = \frac{\kappa E_y}{3\gamma^2}.
\label{eq:ap_pS_scale}
\end{equation}
Furthermore, employing approach similar to simplification of equations \eqref{eq:ap_rtrac} and \eqref{eq:ap_ytrac} to \eqref{eq:ap_rtrac_simp} and \eqref{eq:ap_ytrac_simp} respectively, i.e. scaling the LHS and RHS of the y-component of fluid-substrate inteface traction-balance condition equally, we have,
\begin{equation}
\frac{3\mu\omega\alpha\epsilon_{0}\gamma^2}{\epsilon^2\kappa E_y} = 1 \implies \kappa = \frac{3\mu\omega\alpha\epsilon_{0}\gamma^2}{\epsilon^2 E_y}.
\label{eq:ap_kapp_scale}
\end{equation}
It is interesting to note that substituting the expression for $\kappa$ in the expression for $p_c$ gives,
\begin{equation}
p_c = \frac{\mu\omega\alpha\epsilon_{0}}{\epsilon^2},
\label{eq:ap_pc}
\end{equation}
implying that the solid pressure physically scales, and indeed acts to balance, the applied load at the fluid-substrate interface. 
Furthermore, the scale for $u_y$, and hence $l$, is obtained as,
\begin{equation}
u_y \sim l \sim \kappa L = \frac{3\mu\omega\alpha\epsilon_{0}\gamma^3 R}{\epsilon^{\frac{3}{2}} E_y}.
\label{eq:ap_l_scale}
\end{equation}
Contrasting this with the scale for $l$ for an compressible substrate,
\begin{equation}
l \sim \kappa L = \frac{\mu\omega\alpha\epsilon_{0}\gamma(1+\nu)(1-2\nu)R}{(1-\nu)\epsilon^{\frac{3}{2}} E_y},
\label{eq:ap_l_scale_compr}
\end{equation}
it can be seen that deflection scales smaller by a factor of $\gamma^2$ for an incompressible substrate in comparison to a compressible substrate for the same magnitude of Young's modulus and imposed load, indicating the `stiffening' effect of incompressibility of the substrate for a thin-coating geometry. The scaling analysis presented here is similar to a that in another soft-lubrication study \cite{Rallabandi2017}.\\
To get more insight into the deformation characteristics for incompressible substrates, a dedicated rigorous analysis of incompressible substrates is required, where a major goal would be the reconciliation of solution obtained using constitute relation as given in equation \eqref{eq:ap_incompr_SS} with the solution obtained using constitute relation as given in equation \eqref{eq:ap_linel_SS} under the limiting case of $\nu \rightarrow 0.5$. Such an endevour isn't attempted here.
\begin{figure}[h!]
\centering
\begin{subfigure}[b]{0.495\textwidth}
\centering
\includegraphics[height=5.50cm]{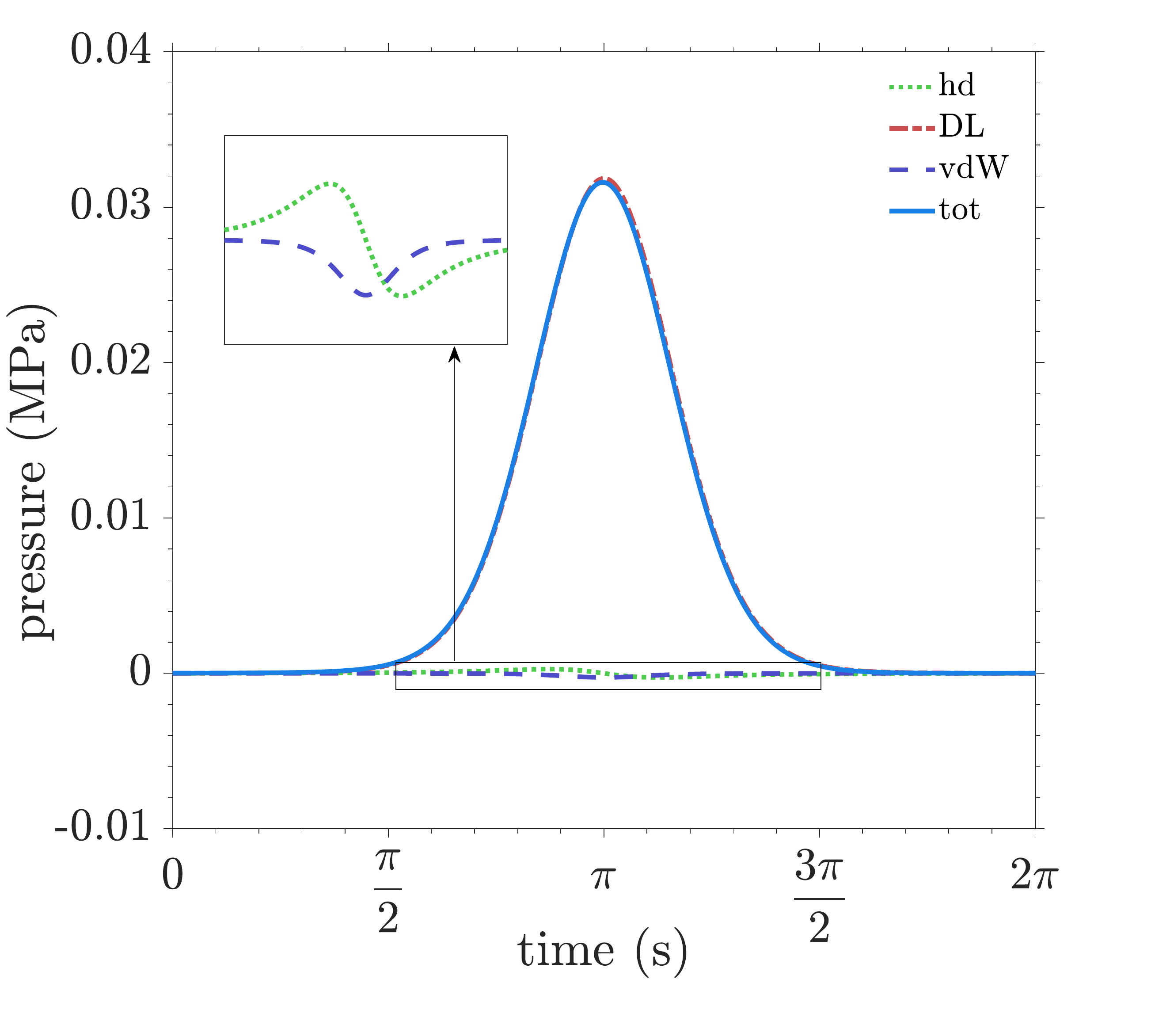}
\caption{\centering}
\label{subfig:ap_pallcl_40d00_rigid}
\end{subfigure}
\begin{subfigure}[b]{0.495\textwidth}
\centering
\includegraphics[height=5.50cm]{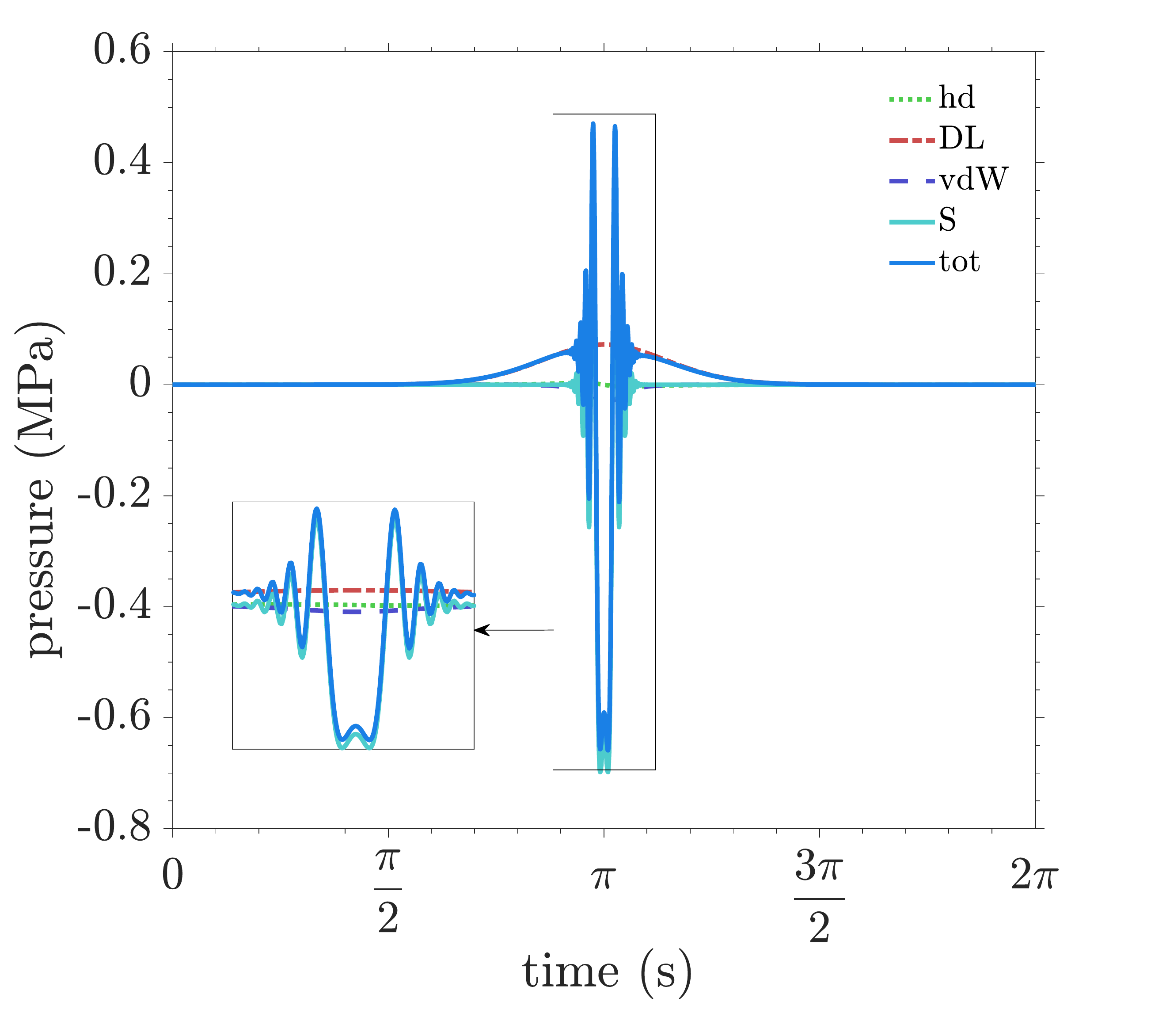}
\caption{\centering}
\label{subfig:ap_pallcl_48d00_rigid}
\end{subfigure}
\begin{subfigure}[b]{0.495\textwidth}
\centering
\includegraphics[height=5.50cm]{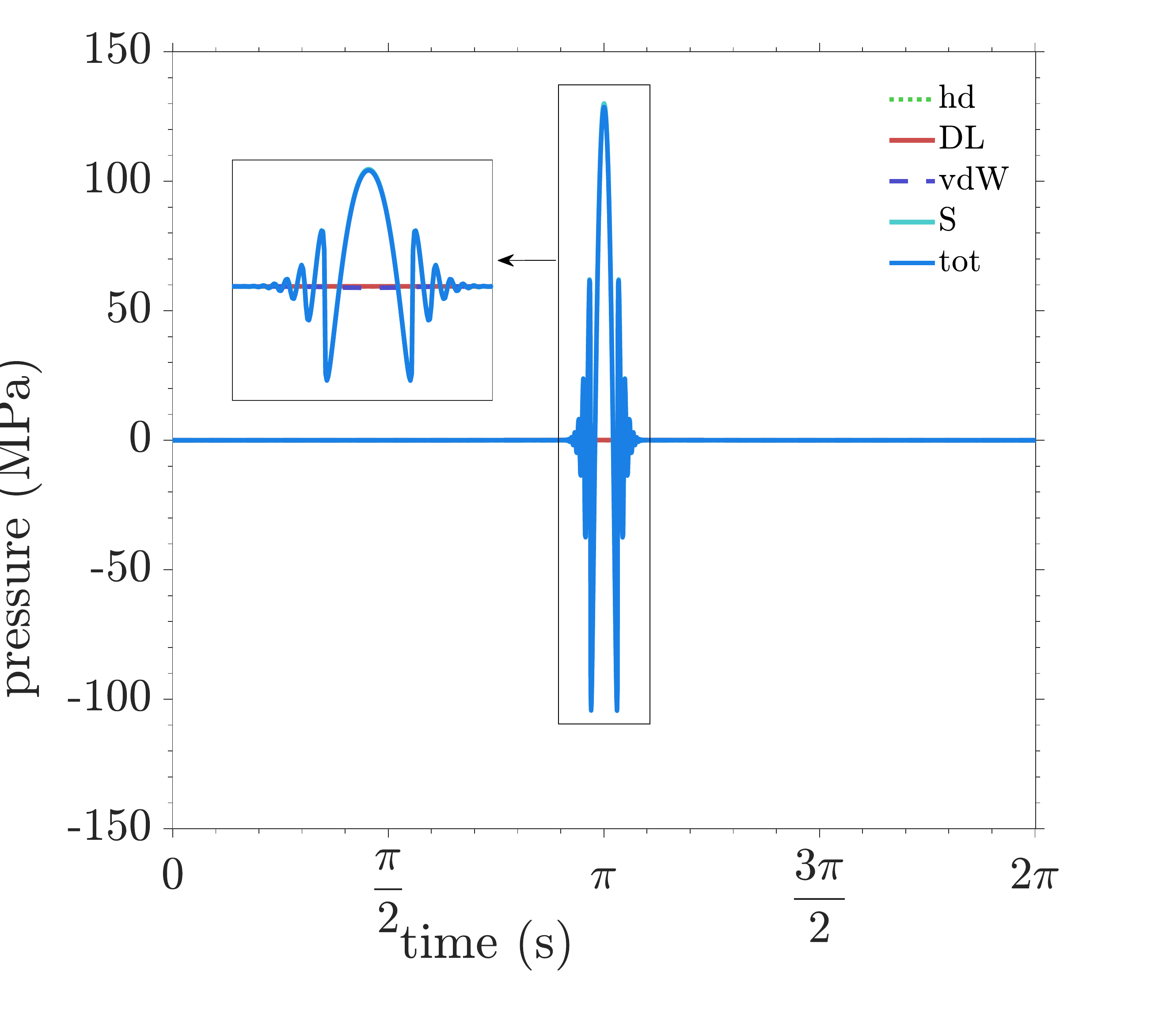}
\caption{\centering}
\label{subfig:ap_pallcl_49d50_rigid}
\end{subfigure}
\caption{Evolution of individual pressure components and total pressure between sphere and substrate at origin with time for full solution for hard substrate for (a) 40 nm amplitude case, (c) 48.0 m amplitude case, and (e) 49.5 nm amplitude case; both the vertical and horizontal axes are linear-scaled}
\label{fig:ap_press}
\end{figure}

\section{Pressure Components}\label{sec:ap_press}
The evolution of individual pressure components as well as total pressure at origin for full solution of hard substrate for the 40 nm, 48 nm and 49.5 nm amplitude cases are presented in Figure \ref{fig:ap_press}. For the 40 nm amplitude case (figure \ref{subfig:ap_pallcl_40d00_rigid}), the EDL disjoining pressure is seen to be dominant and the other pressure components negligible. Furthermore, there is an expected rise in the EDL disjoining (and thus total) pressure near mid-oscillation, which dies down towards $t=0$ as well as $t=2\pi$. The inset shows the van der Waals pressure and hydrodynamic pressure, with their expected trends as per equation \eqref{eq:p0st} and \eqref{eq:pivdW}. Solvation pressure is practically zero and therefore not depicted. For the 48 nm amplitude case (figure \ref{subfig:ap_pallcl_48d00_rigid}), while solvation pressure strongly influences total pressure and leads to rapid and strong fluctuations near mid-oscillation, EDL disjoining pressure can be seen to have a finite value and thus cause a distinct offset of total pressure from solvation pressure. The van der Waals pressure also attains a significant value close to mid-oscillation. Hydrodynamic pressure is negligible throughout. For the 49.5 nm amplitude case (figure \ref{subfig:ap_pallcl_49d50_rigid}), solvation pressure becomes practically the only pressure component and equal to the total pressure.
\section{Solvation Pressure}\label{sec:ap_piS}
For the parameter values corresponding to solvation force for our study, we considered the study by Trokhymchuk et al, 2001 \cite{Wasan2001}, who executed a theoretical study with the aim of obtaining the depletion force associated with structuring of hard-sphere solvents between rigid planar surfaces. Employing Percy-Yevick theory for hard-sphere-like fluid to solve (both asymptotically and computationally) the Orstein-Zernike relations for direct and total correlation functions for two large hard spheres dispersed in a fluid of smaller hard-sphere solvent, they obtained the expressions for depletion force and interaction energy, and the validated the results with multiple MD and theoretical studies. They observed that the decay length and oscillation frequency of the obtained disjoining pressure and interaction energy contributions depended exclusively on fluid's bulk properties i.e. volume fraction and hard-core diameter. Although the term used for the solvent-mediated force in their study is `Depletion Force', it is same as the `Solvation Force' that we consider in our study. We have taken the solvation pressure parameters for our study based on the depletion pressure profile for a solvent of volume fraction 0.3665 (which is close to water's volume fraction, 0.38) obtained by them.\\
While the study by Trokhymchuk et al, 2001 \cite{Wasan2001} is a contemporary detailed exposition into solvation force, experimental and theoretical investigations on force at short-range (i.e. $\sim$ 0.5-2 nm) between surfaces with intervening fluid has been an active area of research for about four decades. From the currently available literature, force between surfaces at small separation can broadly be classified into three components - solvation force (or solvent-structuration force), hydration force, and surface-structuration force \cite{Israelachvili2011}. Solvation force is the excess pressure generated due to oscillations in the packing of solvent molecules in the confinement between surfaces from optimal to pessimal \cite{Israelachvili2011}. This force is typically damped oscillatory in nature and exists for separations upto five to ten times the solvent particle hard-core diameter \cite{Snook1980}. Hydration force is another force due to the interactions of water molecules, but is different from solvation force in the sense that the former is because of hydration effects, i.e. orientational packing and steric hindrance of water molecules hydrated to cations on the surface (in contrast to the latter, which is due to packing efficiency). Hydration force typically appears as an additive to solvation force when computing force at short-range \cite{Pashley1984}. Lastly, for surfaces that are not inert and molecularly smooth, surface features and energy interactions with the fluid lead to alterations to the solvation force (which is typically in the nature of smoothening out of oscillations in the force variation \cite{Qin2003,Yang2011}) as well as to a force due to direct interaction of the surface features. The latter is referred to as structuration force. The exact combination of all such short-range forces for a particular system requires an in-depth analysis of the material properties of media involved and associated energy interactions as well as the structural configuration. However, for molecularly smooth surfaces with intervening fluid as dilute electrolytic solution, the force at short-range retains a damped-oscillatory variation with separation \cite{Pashley1984}. \\
The fundamental groundwork of solvation forces can be traced back to the f study by Asakura and Oosawa, 1954 \cite{Asakura1954}, who studied the osmotic pressure generated between two surfaces in a solution of macromolecules due to depletion of the macromolecules from the intervening region. While the depletion leads to depletion force for such small separations, structuring of molecules between surfaces at somewhat larger separations lead to osmotic forces as well \cite{Trefalt2016}. Similar to the structuring of macromolecules in a solvent leading to small-confinement osmotic force, structuring of solvent in vacuum leads to analogous small-confinement osmotic force. Some of the very first theoretical studies on solvation forces were undertaken in the 1970s \cite[]{Henderson1976,Snook1978,Saam1978,Chan1978}, which were soon well-supported by other experimental studies \cite{Christenson1984,Pashley1984}. Pashley, 1981-1982 \cite{Pashley1981,Pashley1981a,Pashley1982} conducted experimental studies of the force between molecularly smooth mica surfaces in electrolytic solutions, where short-range repulsive hydration forces beyond a critical bulk concentration were observed. Extending on these studies, Pashley and Israelachvili, 1984 \cite{Pashley1984} meausured the force between mica surface in 1 mM aqueous KCl solution focussing on separations below 2 nm, where the observed short-range force profile indicated superposition of a damped-oscillatory force (attributed to solvation effects) and a monotonic exponentially decaying repulsive force (attributed to hydration effects), with the latter being stronger for higher concentrations of KCl. Christenson, 1984 \cite{Christenson1984} measured the force between molecularly smooth mica surfaces immersed in methanol and acetone (H-bonding polar liquids), establishing that oscillatory solvation forces replace continuum van der Waals forces at small separations, much like non-polar fluids. With these studies constituting a groundwork on short-range surface forces, Israelachvili and McGuiggan, 1988 \cite{Israelachvili1988} presented a summary article where they categorized the force between surfaces or particles in liquids into four components - van der Waals force (monotonically attractive), EDL disjoining force (monotonically repulsive), solvation, structural, and hydration forces (monotonic/oscillatory), and repulsive entropic forces (because of thermal motions of protruding surface groups or fluidlike interfaces). Subsequently from the 1990s till now, there have been numerous molecular dynamics, Monte-Carlo and statistical mechanics studies on solvation forces \cite{Snook1980,Kinoshita1996,Gao1997,Gao2007,Yang2011}. Frink and van Swol, 1998\cite{Frink1998} conducted extensive GCMC simulations of LJ fluids between rough walls where they characterized the effects of wall roughness on solvation force. Qin and Fichthorn, 2003 \cite{Qin2003} performed molecular dynamics simulations of solvation and van der Waals forces between nanoparticles in LJ liquid, where damped oscillatory force-separation profile was recovered for solvophilic particles.\\
\end{appendices}

\bibliographystyle{unsrt}
\bibliography{refs_short}

\end{document}